%% file: main.tex
\documentclass[11pt, a4paper]{article}

\usepackage{bbm}
\usepackage[intlimits]{amsmath}
\usepackage{amssymb}
\usepackage{amsfonts}
\usepackage{tikz}
\usetikzlibrary{arrows,cd}
\RequirePackage{snapshot}
\usepackage{setspace}
\usepackage{bm}
\usepackage{graphicx}
\usepackage[a4paper,margin=4cm]{geometry}
\usepackage[numbers]{natbib}
\usepackage[labelfont=bf, font=small, skip=0pt]{caption}
\usepackage{color}
\usepackage{mathrsfs}
\usepackage{dsfont}
\usepackage{esint}
\usepackage{bibentry}
\usepackage{setspace}
\usepackage{arydshln}
\usepackage{cases}
\usepackage{hyperref}
\usepackage{url}
\usepackage{wasysym}
\usepackage{calc}
\usepackage{pgf}
\usepackage{multirow}
\usepackage{lineno}
\usepackage{fancyhdr}
\usepackage{enumitem}
\usepackage[explicit]{titlesec}
\usepackage{subcaption}

\usepackage[utf8]{inputenc}
\usepackage[T1]{fontenc}

\definecolor{BLUE}{rgb}{0.00, 0.00, 0.66}
\definecolor{RED}{rgb}{0.86, 0.00, 0.00}
\definecolor{LRED}{rgb}{0.89, 0.19, 0.19}

\makeatletter
\newcommand{\specialnumber}[1]{%
  \def\tagform@##1{\maketag@@@{(\ignorespaces##1\unskip\@@italiccorr\emph{#1})}}%
}
\newcommand{\specialeqref}[2]{\begingroup
  \def\tagform@##1{\maketag@@@{(\ignorespaces##1\unskip\@@italiccorr\emph{#2})}}%
  \eqref{#1}\endgroup}
\makeatother

\usepackage[singlelinecheck=false,labelfont=it,textfont=normalfont,singlelinecheck=off,justification=raggedright]{subcaption}

\newcommand{\rd}{\,\mathrm{d}}

\newcommand{\od}[2]{\dfrac{\mathrm{d} #1}{\mathrm{d} #2}}

\newcommand{\pd}[2]{\dfrac{\partial #1}{\partial #2}}

\newcommand{\vc}[1]{\bm{#1}}

\newcommand{\articleTitle}[0]{Transport and mixing in control volumes through the lens of probability}

\renewcommand{\subsectionmark}[1]{}
\titleformat{\subsubsection}[runin]{}{\thesubsubsection\quad\caps{#1}.}{1em}{}
\titleformat{\subsection}{\filcenter}{\thesubsection\quad\emph{#1}\indent}{1em}{}
\titleformat{\section}{\filcenter\bf}{\thesection\quad#1}{1em}{}
\titleformat{name=\section,numberless}[block]{\filcenter\bf}{}{1em}{#1}

\begin{document}

\begin{center}
\vspace{-1cm}
\noindent{\LARGE\bf\articleTitle\par}
\vspace{1em}
{\sc John Craske$^{1}$ and Paul Mannix$^{1}$}\\
{\small $^{1}$Department of Civil and Environmental Engineering,\\ Imperial College London, London SW7 2AZ, UK}\\
{\small \today}
\end{center}

\thispagestyle{empty}

\begin{abstract}
  A partial differential equation governing the global evolution
  of the joint probability distribution of an arbitrary number of
  local flow observations, drawn randomly from a control volume,
  is derived and applied to examples involving irreversible
  mixing. Unlike local probability density methods, this work
  adopts a global integral perspective by regarding a control
  volume as the sample space. Doing so enables the divergence theorem to
  be used to expose contributions made by uncertain or stochastic
  boundary fluxes and internal cross-gradient mixing in the
  equation governing the joint probability distribution's
  evolution. Advection and diffusion across the control volume's
  boundary result in source and drift terms, respectively, whereas
  internal mixing, in general, corresponds to the sign-indefinite
  diffusion of probability density. Several typical circumstances
  for which the corresponding diffusion coefficient is negative
  semidefinite are identified and discussed in detail. The
  global joint probability perspective is the natural setting for
  available potential energy and the incorporation of uncertainty
  into bulk, volume integrated, models of transport and
  mixing. Finer-grained information in space can be readily
  obtained by treating coordinate functions as observables. By
  extension, the framework can be applied to networks of
  interacting control volumes of arbitrary size.
\end{abstract}
\section{Introduction}

\subsection{Bulk models and uncertainty}

Bulk, integral, lumped or coarse-grained models in fluid mechanics
often involve integrating equations that govern the local
(pointwise in space) evolution of a system over a control
volume. The divergence theorem \citep{FraTboo2011a} can be used in
their derivation to factorise surface transport terms and internal
dissipative terms.  Such models are useful for providing a
macroscopic picture at scales that are directly relevant to a
given application. However, over a century of intensive research
into fluid turbulence has shown that nature does not always yield
to coarse representations, demanding, in return, case-dependent
closures and probabilistic approaches.

Probabilistic approaches are appropriate here because the heterogeneous contents of control
volumes in bulk models is a source of uncertainty, in addition to local sources of noise or measurement uncertainty in real applications. A relevant
example comes from the field of building ventilation, where it is
often assumed that the air in each room or `zone' of a building is
`well mixed' and, therefore, of uniform temperature. However, it
is now acknowledged, not least due to concerns raised during the
COVID-19 pandemic, that the secondary flows and temperature
structures within rooms play an important role in determining the
fate of contaminants, energy demands and thermal comfort
\citep[see, for example][]{BhaRafm2024a, VouCflo2023a}. An
additional complication, which renders the underlying challenge
probabilistic, is that the occupancy and boundary conditions that
are responsible for producing this heterogeneity are almost never
known precisely \citep{ShaHbae2019a}.

On this basis, the present work forsakes detailed local
deterministic information in physical space for a limited amount
of probabilistic information about an entire control volume. More
precisely, we consider a projection of the complete statistical
(sometimes referred to as `functional') formulation of the
Navier-Stokes equations \cite{HopEpam1948a, LewRpam1962a,
  LunTpof1967a, ComWboo1990a, MonAboo2013a}. The resulting objects
correspond to spatial integrals of local (pointwise) joint
probability distributions and therefore address \emph{what} is
inside the control volume at the expense of knowing precisely
\emph{where} it is occurring.

\subsection{Background}

The full functional formulation of the Navier-Stokes equations
\citep{HopEpam1948a, LewRpam1962a}, is an infinite-dimensional
problem. Notwithstanding its theoretical importance, it is
intractable and unsuitable as an operational approach for
applications. A vast number of studies have therefore sought
closure to the problem by truncating the infinite hierarchy of
moments or cumulants that one can obtain from the functional
equations \citep[see, for example][]{PopSboo2000a, DurKafm2019a}.
On the other hand, a happy consequence of lifting the problem to
an infinite-dimensional space is that the functional formulation
renders the problem linear. Indeed, this idea underpins Koopman
operator theory \citep{KooBpns1931a, MezIafm2013a}, which has
facilitated sophisticated exploration of the infinite-dimensional
problem, with researchers searching for invariant subspaces of the
associated linear operator \citep{BruSpls2016a}. An alternative
perspective, based on the Frobenius-Perron operator, is to
consider the dual problem that focuses on the evolution of
probability distributions \citep{GasPboo1998a, LasAboo1998a}.

In an engineering context, probability density function (PDF)
methods have traditionally focused on local (i.e. conditional on
single or multiple points in space) joint PDFs and were developed
towards the latter half of the twentieth century, primarily in the
fields of combustion \cite{PopSmis1985a, PopSafm1994a,
  PopSboo2000a} and those involving turbulent dispersion
\citep{HunJafm1985a, FisHboo1979a}. The associated governing
equations can be derived from a Lagrangian or Eulerian perspective
but, not typically accounting for the multi-point statistics that
are embedded within the full functional formulation, contain
expectations of gradients that require closure. One attractive
feature of PDF methods is that advection and forcing terms that
can be expressed as functions of the dependent variables appear in
closed form \citep{PopSboo2000a}. Amongst many applications, their
use recently can be found in a derivation of the evolution
equation for the probability density of vorticity from the
Navier-Stokes equations \citep{LiJprs2022a} and analysis of
Rayleigh-B\'{e}nard convection \citep{LulJnjp2011a,
  LulJjfm2015a}.

Our work differs from classical PDF methods
\citep{PopSboo2000a} in focusing on the heterogeneous contents of
an entire control volume rather than point measurements. Unlike
the Dirac measures obtained from sampling a deterministic field at
a fixed point in space and time (for which the question of whether
a given value was observed has a binary answer), PDFs of
deterministic fields sampled over an entire control volume are
more complicated, because they account for the fields' spatial
variability (see figure \ref{fig:lorenz} and the corresponding
discussion in \S\ref{sec:example}). A further way in which this
work differs from previous use of classical PDF methods is in deriving the
governing equations from a dual perspective using the
infinitesimal generator of the Koopman operator. Such an approach
corresponds to the classical derivation of the (weaker) forward
Kolmogorov equation from the (stronger) backward Kolmogorov
equation that is typically presented in textbooks on stochastic
differential equations \cite[e.g.][]{PavGboo2014a}. A
practical advantage in proceeding from a dual perspective is in
making the incorporation of boundary conditions and the
application of the divergence theorem easier to formulate.

Besides being objects that can be evolved in time by
Frobenius-Perron and Fokker-Planck operators, it can hardly be
overstated that PDFs are playing an increasingly prevalent role in
the interpretation of data and development of models from a
Bayesian perspective \citep{BroMjfm2022a, DurKafm2019a}. In
particular, recent work on the spatiotemporal intermittency of
ocean turbulence and its associated mixing employed PDFs to characterise the dissipation of both kinetic and
potential energy \cite{CaeBprl2021a, CouMgrl2021a, LewSjfm2023a},
and motivates the role of spatial variability in the present
work.

An additional advantage of describing the contents of a control volume using PDFs is that several bulk energetic quantities emerge naturally as functionals. In particular, the reference state that is used to define global
available potential energy \citep{MarMmis1903a, LorEtel1955a,
  WinKjfm1995a, TaiRafm2013a}, which quantifies the maximum amount
of potential energy that can be released during a
volume-preserving and adiabatic rearrangement of fluid parcels, is
a functional of the joint PDF of buoyancy and geopotential
height. While such constructions are difficult to wrestle with in
physical space, due to the global nature of the rearrangement,
their expression in terms of joint PDFs is natural and therefore
convenient. In the evolution equation of a sensibly chosen joint
PDF, available potential energy and the vertical buoyancy flux
become known quantities and could, in theory, be used in a
prognostic capacity to model unclosed terms. Whilst it has been
known for some time that the reference state in APE constructions
is closely related to the cumulative distribution of buoyancy over
a domain \cite{TseYphf2001a}, the correspondence does not appear
to have been exploited as a means of modelling stratified fluids
or linked with existing work on local PDF methods.

The present work is aimed at engineers and physicists who wish to
diagnose or model bulk transport and mixing processes from a
probabilistic perspective, rather than mathematicians seeking to
obtain rigorous results from the Navier-Stokes
equations. Accordingly, we will often assume existence and
sufficient smoothness of the objects being manipulated without
stating so explicitly. The resulting equations are nevertheless of
mathematical interest and perhaps worthy of study in their own
right. They are effectively a projection of the full functional
formulation of the Navier-Stokes equations \citep[see, for
example,][]{ComWboo1990a} and therefore contain conditional
expectations that require closure. If such terms are regarded as a
function of the projected system's joint PDF, the resulting
partial differential equation is a nonlinear Fokker-Planck
equation reminiscent of mean field theory \citep{FraTboo2005a,
  BarVboo2024a} and the associated McKean-Vlasov stochastic
differential equation \cite{McKHnas1966a}. The approach might
therefore provide a means of generating operational stochastic
models or, at least, provide an alternative means of interpreting
existing low-dimensional stochastic models of turbulence
\citep[see, for example,][]{KerAjfm1999a, WunSjfm2005a}. A further
aspect of this work that might elicit broader interest is its
connection with renewed interest in Koopman von Neumann mechanics
to cast nonlinear dynamical systems into a form that is amenable
to quantum computation \citep{LinYarx2022a}.

The derivation of the governing evolution equation for probability
distributions of control volumes is presented in \S\ref{sec:gov} and is followed by a discussion of the associated diffusion coefficient in \S\ref{sec:D2}. To illustrate application of the approach, several examples  are then discussed in \S\ref{sec:examples}. First, we illustrate the basic ideas behind
viewing the contents of a control volume probabilistically, with
relatively simple introductory examples for one- and
two-dimensional domains.

\subsection{Introductory examples}
\label{sec:example}

This work addresses the following question: \emph{if a point
  $\omega$ belonging to a spatial domain or control volume
  $\Omega$ is selected at random, what is the probability that a
  given vector of field variables $\vc{Y}$, evaluated at that point at a particular time, will have values lying in a given range?}

Here, the term `\emph{at random}' refers to the usual Lebesgue
measure for space, which means that the probability of selecting a
point within a given volume is proportional to the physical size
of the volume. An Eulerian \emph{point} $\omega$ becomes an
element of a sample space $\Omega\ni\omega$ parameterised by
coordinate functions $\vc{X}$. The field variables $\vc{Y}=\varphi(\vc{X})$ are
quantities such as velocity, temperature or scalar
concentration. We are therefore interested in the distribution of
$\vc{Y}$ across the entire control volume.

The answer to the question is given by the probability density
(or, more generally, distribution) associated with $\vc{Y}$, which
does not contain information about the precise relationship
between $\vc{Y}$ and $\vc{X}$. Indeed, the probability density
corresponds to an infinite number of possible functions $\varphi$
of $\vc{X}$ that produce the same distribution of $\vc{Y}$ over
the control volume. Informally, the construction involves tipping
the fields $\vc{Y}$ into a sack that doesn't store values of
$\vc{X}$; the probability density `weighs' the various values of
$\vc{Y}$, without caring about where they came from.

\begin{figure}
  \input{example1d}
  \vspace*{1mm}
\caption{The probability distribution $f_{Y}$ (right) of the value of a function $Y=\varphi(X)$ (left)
  parameterised by the coordinate $X$. Stationary points of the
  function $Y$ (highlighted by horizontal lines) correspond to
  singularities in the distribution $f_{Y}$. To account for parts of
  $Y$ that are constant (dashed), the distribution contains
  a Dirac measure $\delta$, weighted by the proportion $\mu$ of
  the domain over which $Y$ is constant.}
\label{fig:example1d}
\end{figure}
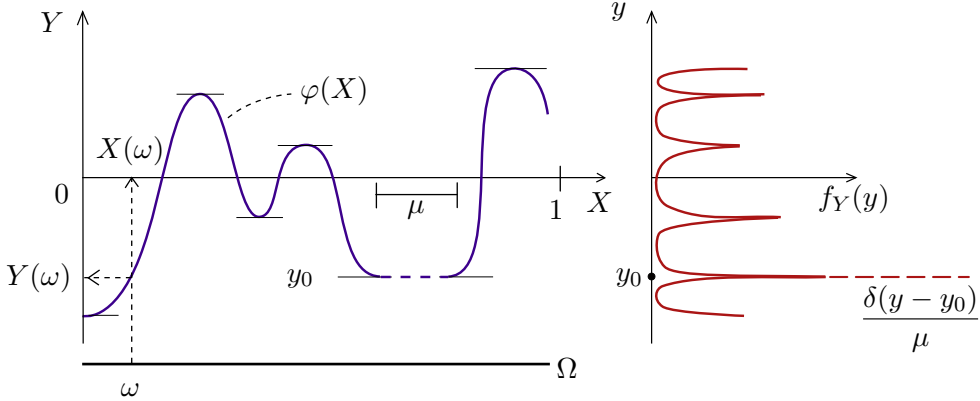

Figure \ref{fig:example1d} illustrates an example in one dimension
for which `$\vc{Y}$' can therefore be replaced with `$Y$'. As explained more precisely
in \S\ref{sec:pushforward}, the construction of $f_{Y}$ involves
determining the proportion of $\Omega$ taken up by a given value
of $Y$. Analytically, this involves considering intervals of
$\Omega$ over which $Y=\varphi(X)$ is strictly monotone with respect to $X$
and, therefore, invertible. Such intervals lie between the
stationary points of $\varphi$. Each point within the interval
contributes a density that is inversely proportional to the
gradient of $\varphi$ with respect to $X$ at that point, because
relatively large gradients in $\varphi$ account for a relatively small
proportion of $\Omega$.

\begin{figure}[t]
  \centering
  \includegraphics[scale=0.9]{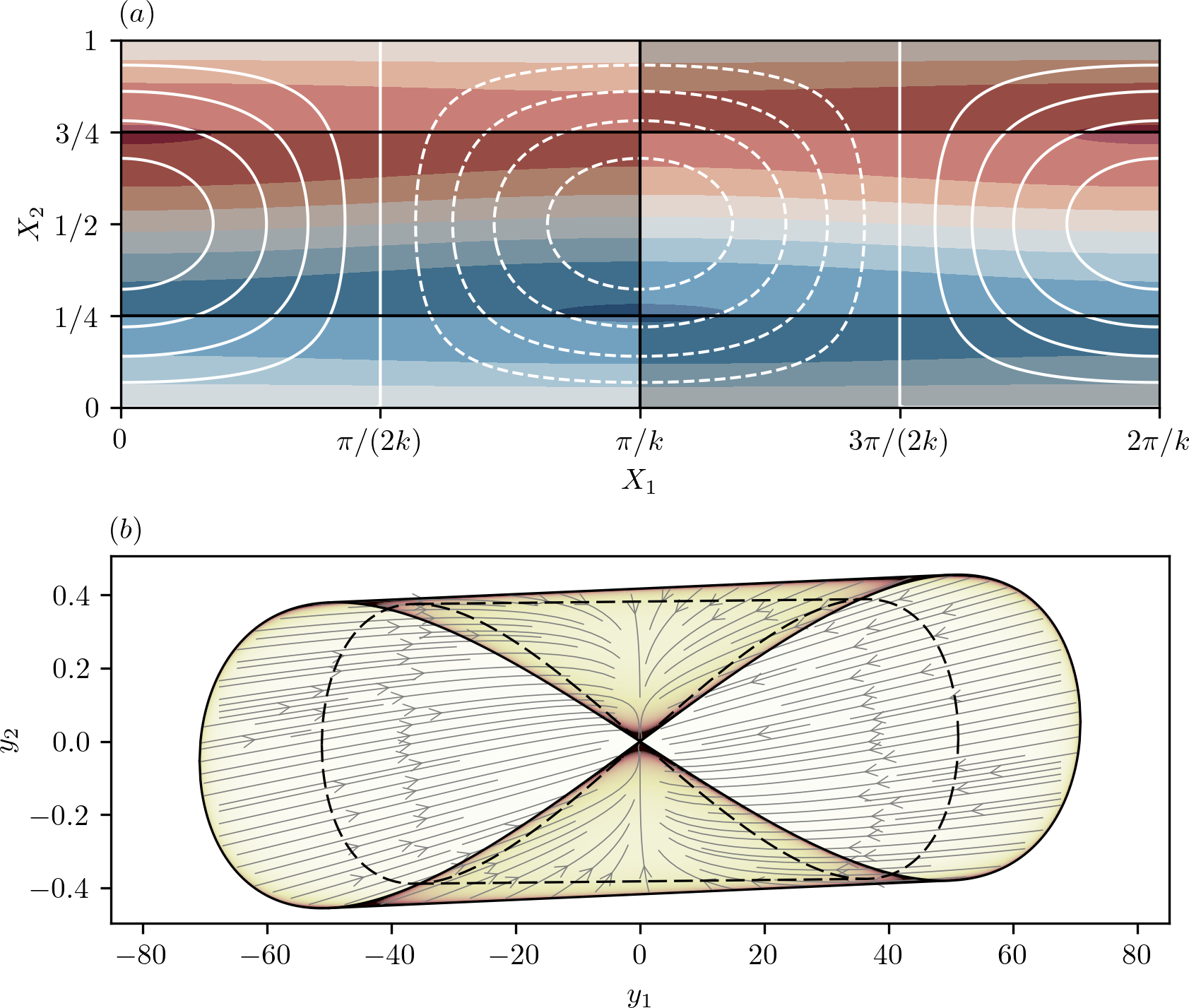}
  \vspace*{1mm}
  \caption{$(a)$ Relative buoyancy (blue to red) and vertical
    velocity isolines (dashed for negative velocity) at time $t$
    of the two-dimensional fields corresponding to a point on the
    Lorenz attractor described in \S\ref{sec:lorenz} ($r=28$,
    $s=10$, $b:=4\pi^{2}(k^{2}+\pi^{2})^{-1}=8/3$ for horizontal
    wave number $k$). $(b)$ The joint probability density (shaded
    colour) of vertical velocity ($y_{1}$) and relative buoyancy
    ($y_{2}$) corresponding to the field shown in $(a)$. The solid
    black line marks singularities in the density and the dashed
    black line corresponds to the position of the singularities at
    $t+0.04$. The light grey arrows are tangential to the
    probability flux induced by the Lorenz equations
    \eqref{eq:lorenz} that is responsible for the time evolution
    of the density.}
  \label{fig:lorenz}
\end{figure}

All intervals in $X$ over which $\varphi$ is invertible contribute
to the density in proportion to their size, which gives the
density the appearance of consisting of folds and caustics (see
also figure \ref{fig:lorenz}b). Whilst values of $Y$ associated
with large gradients contribute relatively small amounts to the
probability density within each interval, their overall
contribution also accounts for the number of intervals in which
they are found across the entire domain. For example, the
probability density of $Y=\varphi(X)=\mathrm{cos}(\pi nX)$ for
$n\in \mathbb{N}^{+}$ and $X\in[0,1)$ is
$f_Y(y)=\pi^{-1}(1-y^{2})^{-1/2}$ for $y\in(-1,1)$, and is
therefore independent of $n$.

Points at which $Y$ is stationary produce singularities in the
distribution $f_{Y}$ shown on the right of figure
\ref{fig:example1d}. In exceptional cases, intervals of finite
size over which $Y=y$ is constant cannot be described by a density
\emph{function} but rather a Dirac measure or distribution,
weighted by the proportion of the domain over which $Y$ is
constant, as indicated by the horizontal dashed part of $\varphi(X)$ and
corresponding $\delta$ in figure \ref{fig:example1d} \citep[see
also chapter 1 of][]{KaiCboo2001a}.

We now consider a two-dimensional example using velocity and
buoyancy fields from Lorenz's 1963 model for convection
\citep{LorEjas1963a} shown in figure \ref{fig:lorenz}a (details of
the model and calculations required to construct the corresponding
distribution $f_{\vc{Y}}$ can be found in \S\ref{sec:lorenz}). In
this example $\vc{X}:=(X_{1},X_{2})^{\top}\in\mathbb{R}^{2}$ are
the horizontal and vertical coordinates and
$\vc{Y}_{t}:=(Y_{t}^{1},Y_{t}^{2})^{\top}=\varphi_{t}(\vc{X})\in\mathbb{R}^{2}$
denotes the vertical velocity and buoyancy, relative to the static state of linear conduction, at time
$t$, respectively. Figure \ref{fig:lorenz}b depicts the
joint probability density
$f_{\vc{Y}}(-,t):\mathbb{R}^{2}\rightarrow \mathbb{R}^{+}$
corresponding to the fields shown in figure \ref{fig:lorenz}a,
such that the probability of finding a value of $\vc{Y}_{t}$ in
any range/codomain $C\subset\mathbb{R}^{2}$ is

\begin{equation}
  \mathbb{P}\{\vc{Y}_{t}\in C\} = \int\limits_{C}f_{\vc{Y}}(\vc{y},t)\rd \vc{y},
\end{equation}

\noindent where $\vc{y}:=(y_{1},y_{2})^{\top}$ denotes the
argument to the probability density corresponding to the
capitalised random variable $\vc{Y}_{t}$. Again, it is important
to appreciate that whilst the density $f_{\vc{Y}}$ determines the volume
average of any observable $g$:

\begin{equation}
  \mathbb{E}[g(\vc{Y}_{t})] = \int\limits_{\mathbb{R}^{2}}g(\vc{y})f_{\vc{Y}}(\vc{y},t)\rd \vc{y},
\end{equation}

\noindent it does not provide information about how
$\vc{Y}_{t}$ is correlated with $\vc{X}$. In particular,
$f_{\vc{Y}}$ does not provide information about multipoint statistics
or spatial gradients, unless they are included in $\vc{Y}_{t}$.

To understand the PDF shown in figure \ref{fig:lorenz}b, it is
helpful to consider the regions over which $\varphi_{t}$ is
invertible. Such regions are highlighted as shaded rectangles in
figure \ref{fig:lorenz}a, in which the Jacobian
$J:=\partial \varphi_{t}/\partial{\vc{X}}$ does not vanish
($|J|\neq 0$). The solid black lines, separating the regions,
denote points for which $|J|=0$, which account for the
singularities in figure \ref{fig:lorenz}b. In particular, $|J|=0$
over the sets $S_{1}:=\{X_{1}=\pi/k, X_{2}\in [0,1]\}$ and
$S_{2}:=\{X_{1}=[0,2\pi/k], X_{2}=1/2\pm 1/4\}$, where $k$ is a
horizontal wavenumber. The set $S_{1}$ corresponds to the solid
black line that looks like `$\infty$' in figure \ref{fig:lorenz}b,
while $S_{2}$ corresponds to the nearly-horizontal lines that
define its convex hull.  As explained for the previous
one-dimensional example, the folded appearance of $f_{\vc{Y}}$
shown in \ref{fig:lorenz}b is due to the fact that a given value
of $\vc{Y}_{t}$ contributes to $f_{\vc{Y}}$ from more than one
region in the domain.

As the fields shown in figure \ref{fig:lorenz}a evolve in
time, the density is transported over the phase space shown in
figure \ref{fig:lorenz}b. Since the density integrates to
unity, it is useful to regard probability, like mass, as a
conserved quantity. From this perspective, the governing equations
for $\vc{Y}_{t}$ over the entire control volume correspond to a
two-dimensional velocity field that produces a flux of density in
phase space. The direction of the density flux for this
example is shown in figure \ref{fig:lorenz}b with grey
lines. The density flux determines the subsequent evolution of
$f_{\vc{Y}}$, whose singularities at short time after $t$ are depicted with
dashed lines in figure \ref{fig:lorenz}b. Understanding how
the evolution of $f_{\vc{Y}}$ depends on the evolution of the field
variables $\vc{Y}_{t}$, particularly in terms of the physical
boundary conditions imposed on $\vc{Y}_{t}$, is the central topic
of this article and addresses the \emph{`at a particular time'} condition of the question posed at the start of this subsection.

\section{Volumetric evolution equations}
\label{sec:gov}

\subsection{Local governing equations}
\begin{figure}
  \begin{center}
    \input{diag}
  \end{center}
  \caption{The sample space $\Omega$ as the domain of random
    variables corresponding to `coordinates' $\vc{X}$ and field
    variables $\vc{Y}_{t}=\varphi_{t}(\vc{X})$. The density $f_{\vc{Y}}$ for $\vc{Y}_{t}$ is
    made narrower by diffusion and subjected to internal forcing terms due to
    advective boundary fluxes of $Y_{t}^{1}$ and $Y_{t}^{2}$ (see labels `in' and `out').}
  \label{fig:diag}
\end{figure}
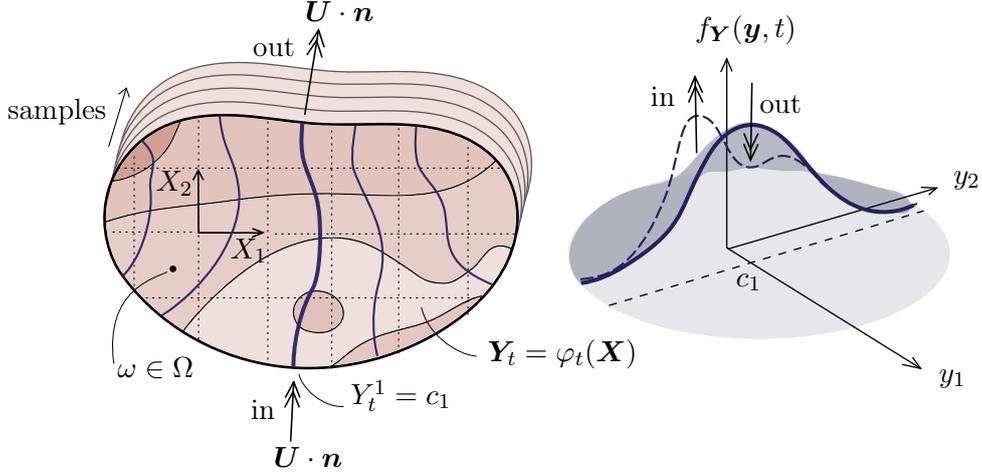

For the purposes of emphasising a probabilistic perspective
that focuses on control volumes, regard the spatial domain
$\Omega$ as a sample space,
$\vc{X}:\Omega\rightarrow \mathcal{X}\subset\mathbb{R}^{d}$ as
coordinate functions and
$\vc{Y}_{t}:\Omega\rightarrow \mathcal{Y}\subset\mathbb{R}^{n}$ as
`random' variables, as illustrated in figure \ref{fig:diag}. Let the
Eulerian evolution of $\vc{Y}_{t}$ at a given point in the domain
be determined by the differential equation

\begin{equation}
  \od{\vc{Y}_{t}}{t}=\vc{Q}_{t}-\vc{U}_{t}\cdot\nabla\vc{Y}_{t}+\vc{\alpha}\Delta\vc{Y}_{t},
\label{eq:gov}
\end{equation}

\noindent where $\vc{Q}_{t}$ represents forcing terms,
$\vc{U}_{t}$ is a solenoidal velocity field
(i.e. $\nabla\cdot\vc{U}_{t}\equiv 0$) and
$\vc{\alpha}\in\mathbb{R}^{n\times n}$ is a (typically diagonal)
matrix containing the diffusivities associated with each component
of $\vc{Y}_{t}$. The framework readily accommodates stochastic
forcing in \eqref{eq:gov}, however, its inclusion does not
directly affect the arguments below and is therefore omitted for
clarity.

Equation \eqref{eq:gov} is cast as an ordinary differential
equation because the relationship $\vc{Y}_{t}=\varphi_{t}(\vc{X})$
is not necessarily known. With $\Omega$ regarded as a sample
space, $\nabla\vc{Y}_{t}$ and $\Delta \vc{Y}_{t}$ will be treated
as symbols denoting (unknown) random variables, rather than known
derivatives of $\varphi_{t}(\vc{X})$. Indeed, without explicit knowledge of
the correlation between $\vc{Y}_{t}$ and $\vc{X}$, spatial
derivatives are not available in this setting, unless they are
included in $\vc{Y}_{t}$. The system \eqref{eq:gov} is therefore
unclosed because it consists of more unknowns than equations. If
spatial derivatives were included in $\vc{Y}_{t}$, their evolution
equations would include higher derivatives and one would be faced
with the infinite hierarchy of equations that constitutes the
closure problem of turbulence \citep{ComWboo1990a}.

To determine the evolution of the probability distribution of $\vc{Y}_{t}$, consider the infinitesimal generator $\mathscr{L}$ acting on an observable $g:\mathbb{R}^{n}\rightarrow \mathbb{R}$:

\begin{equation}
\mathscr{L}g(\vc{y}):=\lim\limits_{t\rightarrow s}\frac{\mathbb{E}_{\vc{Y}_{s}}[g(\vc{Y}_{t})]-g(\vc{y})}{t-s}=\mathbb{E}_{\vc{Y}_{s}}\left[\pd{g(\vc{Y}_{s})}{Y_{s}^{i}}\od{Y^{i}_{s}}{t}\right],
\label{eq:L}
\end{equation}

\noindent where the conditional expectation
$\mathbb{E}_{\vc{Y}_{s}}[\cdot]:=\mathbb{E}[\cdot|\vc{Y}_{s}=\vc{y}]$
accounts for the behaviour of all points within the domain where
$\vc{Y}_{s}=\vc{y}$. The generator can therefore be
expressed in terms of \eqref{eq:gov} as

\begin{equation}
  \mathscr{L}g(\vc{y},s)=\mathbb{E}_{\vc{Y}_{s}}\left[(\vc{Q}_{s}^{i}-\vc{U}_{s}\cdot\nabla Y^{i}_{s}+\alpha_{ij}\Delta Y_{s}^{j})\pd{g(\vc{Y}_{s})}{Y_{s}^{i}}\right].
\label{eq:generator}
\end{equation}

We will follow standard arguments to construct forward and
backward Kolmogorov equations in terms of the generator
$\mathscr{L}$ \citep[see, e.g.][]{GarCboo1989a,
  PavGboo2014a}. Along the way, the divergence theorem will be applied to
\eqref{eq:gov} to produce boundary and irreversible mixing terms
that will appear in the associated Kolmogorov equations as forcing
and anti-diffusion terms, respectively.

\subsection{The divergence theorem}
\label{sec:stokes}

To understand the physics behind the evolution of probability
density over a control volume, it is useful to decompose the
transport terms in \eqref{eq:generator} into internal terms,
associated with irreversible mixing, and transport across the
control volume's boundary. To do so, it is necessary to recognise
the connection between global expectations obtained via
conditioning on $\vc{X}$ compared with conditioning on
$\vc{Y}_{s}$. In this regard, recall the elementary property of
conditional expectations:
\begin{equation}
\mathbb{E}=\mathbb{E}\circ\mathbb{E}_{\vc{Y}_{s}}
=\mathbb{E}\circ\mathbb{E}_{\vc{X}}
\label{eq:EE}
\end{equation}

\noindent which means that global expectations can be computed
from expectations conditioned on $\vc{Y}_{s}$ or expectations
conditioned on $\vc{X}$. The latter are useful because they
correspond to variables in physical space to which it is possible
to apply the divergence theorem in order to extract boundary
fluxes. First, consider the advective transport
$\vc{U}_{t}\cdot\nabla \vc{Y}_{t}$, using
$\nabla\cdot\vc{U}_{t}\equiv 0$ and the chain rule:

\begin{equation}
  \mathbb{E}_{\vc{Y}_{s}}\left[\vc{U}_{s}\cdot(\nabla Y_{s}^{i})\pd{g}{Y_{s}^{i}}\right]
= \mathbb{E}_{\vc{Y}_{s}}[\nabla\cdot (\vc{U}_{s}g)].
\end{equation}

\noindent Next, apply \eqref{eq:EE} to deduce that

\begin{equation}
\mathbb{E}[\mathbb{E}_{\vc{Y}_{s}}[\nabla\cdot (\vc{U}_{s}g)]]
=\mathbb{E}[\mathbb{E}_{\vc{X}}[\nabla\cdot (\vc{U}_{s}g)]],
\label{eq:surf}
\end{equation}

\noindent which is a normalised volume integral of a divergence
and can therefore be converted into a surface integral:

\begin{equation}
\mathbb{E}[\mathbb{E}_{\vc{X}}[\nabla\cdot (\vc{U}_{s}g)]]=\frac{1}{\mu^{d}(\mathcal{X})}\int\limits_{\partial\mathcal{X}}\mathbb{E}_{\vc{X}}[\vc{n}\cdot\vc{U}_{s}g]\rd \vc{x},
\label{eq:Ediv}
\end{equation}

\noindent where $\partial\mathcal{X}$ denotes the domain's
boundary with outward unit normal $\vc{n}$ and
$\mu^{d}(\mathcal{X})$ is the $d$-dimensional size of the
domain's volume.

Let $\vc{Y}_{s}$ have a probability density $f_{\vc{Y}}$, and let
$f_{\vc{Y}|\partial\Omega}$ correspond to the density that is
conditional on $\vc{Y}_{t}$ being sampled from the boundary of the
domain. Assuming that $f_{\vc{Y}|\partial\Omega}\neq 0$ implies
that $f_{\vc{Y}}\neq 0$ (or, more technically, that the measure
associated with $f_{\vc{Y}|\partial\Omega}$ is absolutely
continuous with respect to that associated with $f_{\vc{Y}}$), the
expected value of $\vc{n}\cdot\vc{U}_{s}g$ on the boundary is 

\begin{equation}
\int\limits_{\mathcal{Y}}\mathbb{E}_{\vc{Y}_{s}|\partial\Omega}[\vc{n}\cdot\vc{U}_{s}g]f_{\vc{Y}|\partial\Omega}\rd \vc{y}
=
  \mathbb{E}\left[\mathbb{E}_{\vc{Y}_{s}|\partial\Omega}[\vc{n}\cdot\vc{U}_{s}g]\frac{f_{\vc{Y}|\partial\Omega}}{f_{\vc{Y}}}\right],
  \label{eq:Esurface}
\end{equation}

\begin{figure}
  \input{shapes}
  \caption{The ratio $\phi$, which quantifies the size of a
    domain's bounding surface relative to the size $\mu^{d}(\mathcal{X})$
    of its interior (see discussion following
    \eqref{eq:Ediv}). The dimension of the interior is indicated
    at the top of the figure. The first three domains are
    $d$-balls enclosed by $(d-1)$-spheres.}
    \label{fig:ratios}
\end{figure}
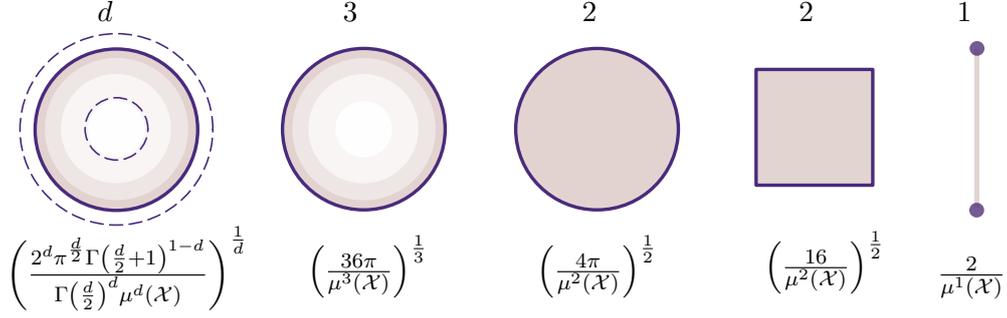

\noindent where $\mathbb{E}_{\vc{Y}_{s}|\partial\Omega}$ is an
expectation conditioned on the boundary. A detailed discussion
about $f_{\vc{Y}|\partial\Omega}$ can be found at the end of
\S\ref{sec:bou}, with a simple example illustrated in figure
\ref{fig:leibniz}.

Noting that \eqref{eq:Ediv} is a volume average, whereas
\eqref{eq:Esurface} is a surface average, their ratio is equal to
the ratio of the domain's surface area to the domain's volume:
\begin{equation}
  \phi:=\frac{\mu^{d-1}(\partial\mathcal{X})}{\mu^{d}(\mathcal{X})}.
  \label{eq:phi}
\end{equation}
\noindent As illustrated alongside various other examples in figure \ref{fig:ratios}, for a closed one-dimensional domain $\mu^{d-1}(\partial\mathcal{X})=1+1$ and, therefore $\phi=2/\mu^{1}(\mathcal{X})$. 

Disaggregating \eqref{eq:surf} into conditionals with respect to
$\vc{Y}_{s}$, on the grounds that the equality holds for all
distributions of $\vc{Y}_{s}$, and using $\phi$ to relate
\eqref{eq:Ediv} and \eqref{eq:Esurface}, implies that

\begin{equation}
\mathbb{E}_{\vc{Y}_{s}}\left[\vc{U}_{s}\cdot\nabla Y^{i}\pd{g}{Y_{s}^{i}}\right]=\phi\,\mathbb{E}_{\vc{Y}_{s}|\partial\Omega}[\vc{n}\cdot\vc{U}_{s}g]\frac{f_{\vc{Y}|\partial\Omega}}{f_{\vc{Y}}}.
\label{eq:stokes}
\end{equation}

Equation \eqref{eq:stokes} confirms the intuitive idea that the
expected evolution of $g$ due to advection by a solenoidal
velocity field is affected only by fluxes of $g$ through the
boundary. The ratio $\phi$
multiplied by $f_{\vc{Y}|\partial\Omega}/f_{\vc{Y}}$ accounts for
the size of the boundary relative to the size of the interior for
a given $\vc{Y}_{s}=\vc{y}$. In particular, noting that
$\mathbb{E}[f_{\vc{Y}|\partial\Omega}/f_{\vc{Y}}]=1$, a uniform
unit flux ($\vc{n}\cdot\vc{U}_{s}g\equiv 1$) out of the
domain causes a reduction in $\mathbb{E}[g]$ at a rate that is
equal to $\phi$.

The decomposition of $\alpha_{ij}\Delta Y_{s}^{j}\partial_{i}g$ in
\eqref{eq:generator} is similar to the decomposition of
$\vc{U}\cdot\nabla\vc{Y}\partial_{i}g$ described above, except for
the fact that commutation of $\nabla\cdot$ and $\partial_{i}g$
produces two terms:
\begin{equation}
    \mathbb{E}_{\vc{Y}_{s}}\left[\Delta Y_{s}^{j}\pd{g}{Y_{s}^{i}}\right]=-\mathbb{E}_{\vc{Y}_{s}}\left[\nabla Y_{s}^{j}\cdot \nabla Y_{s}^{k}\frac{\partial^{2}g}{\partial Y_{s}^{i}\partial Y_{s}^{k}}\right]+\mathbb{E}_{\vc{Y}_{s}}\left[\nabla\cdot\left(\nabla Y_{s}^{j}\pd{g}{Y_{s}^{i}}\right)\right].
\label{eq:EE3}
\end{equation}

\noindent Application of the steps \eqref{eq:EE}-\eqref{eq:stokes} above
to the final term, which corresponds to a volume integral of a divergence of a flux, shows that 

\begin{equation}
  \mathbb{E}_{\vc{Y}_{s}}\left[\nabla\cdot\left(\nabla Y_{s}^{j}\pd{g}{Y_{s}^{i}}\right)\right]=
  \phi\,\mathbb{E}_{\vc{Y}_{s}|\partial\Omega}\left[\vc{n}\cdot\left(\nabla Y_{s}^{j}\pd{g}{Y_{s}^{i}}\right)  \right]\frac{f_{\vc{Y}|\partial\Omega}}{f_{\vc{Y}}}.
\label{eq:stokes2}
\end{equation}

\noindent The right-hand side of \eqref{eq:EE3} therefore
decomposes the effects of diffusion into a term that describes
irreversible internal mixing/dissipation (first term) and boundary
fluxes given by \eqref{eq:stokes2} (second term).

\subsection{The backward Kolmogorov equation}
\label{sec:backward}

For random variables $\vc{Y}'_{t}$ and $\vc{Y}_{t}$, and a function
$g:\mathbb{R}^{n}\rightarrow\mathbb{R}$, the expected
value of $g(\vc{Y}_{t})$, conditional on $\vc{Y}_{t}=\vc{y}$,
is $g(\vc{y})$. Therefore
$\mathbb{E}_{\vc{Y}_{t}}[\vc{Y}'_{t}g(\vc{Y}_{t})]=\mathbb{E}_{\vc{Y}_{t}}[\vc{Y}'_{t}]g(\vc{y})$ and 
\begin{equation}
  \mathscr{L}g(\vc{y})=\mathbb{D}^{(0)}g+\mathbb{D}^{(1)}_{i}\pd{g}{y_{i}}+\mathbb{D}^{(2)}_{ij}\frac{\partial^{2}g}{\partial y_{i}\partial y_{j}}.
\end{equation}
\noindent The coefficient $\mathbb{D}^{(0)}$ of the `source term' corresponds to
\eqref{eq:stokes} and is due to fluid that is advected into and
out of the domain:

\begin{equation}
\mathbb{D}^{(0)}:=-\phi\,\mathbb{E}_{\vc{Y}_{s}}[\vc{n}\cdot\vc{U}_{s}]\frac{f_{\vc{Y}|\partial\Omega}}{f_{\vc{Y}}}.
\end{equation}

\noindent Using \eqref{eq:stokes2}, the so-called drift velocity is

\begin{equation}
\mathbb{D}^{(1)}_{i}:=\mathbb{E}_{\vc{Y}_{s}}[Q_{s}^{i}]+\phi\,\mathbb{E}_{\vc{Y}_{s}|\partial\Omega}\left[\alpha_{ij}\vc{n}\cdot\left(\nabla Y_{s}^{j}\right)  \right]\frac{f_{\vc{Y}|\partial\Omega}}{f_{\vc{Y}}},
\label{eq:D1}
\end{equation}

\noindent and the symmetric diffusion coefficient is

\begin{equation}
\mathbb{D}^{(2)}:=-\frac{1}{2}\mathbb{E}_{\vc{Y}_{s}}\left[\vc{\alpha}\nabla \vc{Y}_{s}\nabla \vc{Y}_{s}^{\top}+\nabla \vc{Y}_{s}\nabla \vc{Y}_{s}^{\top}\vc{\alpha}^{\top}\right].
\label{eq:D2}
\end{equation}

\noindent If an explicit stochastic term in the form of a Wiener
process were added to \eqref{eq:gov}, it is readily shown that
$\mathbb{D}^{(2)}$ would include a positive semidefinite diffusion
matrix \cite{PavGboo2014a}.

If $v(\vc{y},s):=\mathbb{E}_{\vc{Y}_{s}}[g(\vc{Y}_{t})]$ then 

\begin{equation}
  -\partial_{s}v(\vc{y},s)=\mathscr{L}v(\vc{y},s),
  \label{eq:backward}
\end{equation}

\noindent is solved backwards in time from the end condition
$v(\vc{y},t)=g(\vc{y})$, so that $v(\vc{y},s)$ is the expected
value of $g(\vc{Y}_{t})$ given $\vc{Y}_{s}=\vc{y}$ for $s<t$. 
The generator $\mathscr{L}$ therefore `pulls back' the observation $g$
along the dynamics specified by \eqref{eq:gov}.

\subsection{The forward Kolmogorov equation}

To derive the so-called `forward' equation corresponding to
\eqref{eq:backward} \citep[see][for further details]{PavGboo2014a}, the time derivative of the
observable's expectation is expressed in terms of the probability distribution $f_{\vc{Y}}$:
\begin{equation}
\partial_{t}\mathbb{E}[g(\vc{Y}_{t})|\vc{Y}_{s}=\vc{y}]=\partial_{t}\int\limits_{\mathcal{Y}}g(\vc{y})f_{\vc{Y}}(\vc{y},t)\rd\vc{y}=\int\limits_{\mathcal{Y}}g(\vc{y})\partial_{t}f_{\vc{Y}}(\vc{y},t)\rd\vc{y}.
\label{eq:dEdt_1}
\end{equation}
\noindent Alternatively, using $\mathscr{L}$ and integrating by parts,
\begin{equation}
\partial_{t}\mathbb{E}[g(\vc{Y}_{t})|\vc{Y}_{s}=\vc{y}]=\int\limits_{\mathcal{Y}}\mathscr{L}g(\vc{y})f_{\vc{Y}}(\vc{y},t)\rd \vc{y}=\int\limits_{\mathcal{Y}}g(\vc{y})\mathscr{L}^{\dagger}f_{\vc{Y}}(\vc{y},t)\rd \vc{y}.
\label{eq:dEdt_2}
\end{equation}
\noindent Therefore, given that \eqref{eq:dEdt_1} and \eqref{eq:dEdt_2} are 
valid for all suitable observables $g$,
\begin{equation}
  \partial_{t}f_{\vc{Y}}(\vc{y},t)=\mathscr{L}^{\dagger}f_{\vc{Y}}(\vc{y},t),
  \label{eq:forward}
\end{equation}
\noindent where 
\begin{equation}
  \mathscr{L}^{\dagger}f_{\vc{Y}}(\vc{y},t):=\mathbb{D}^{(0)}f_{\vc{Y}}(\vc{y},t)-\pd{}{y_{i}}(\mathbb{D}^{(1)}_{i}f_{\vc{Y}}(\vc{y},t))+\frac{\partial^{2}}{\partial y_{i}\partial y_{j}}(\mathbb{D}_{ij}^{(2)}f_{\vc{Y}}(\vc{y},t)),
\label{eq:kf}
\end{equation}
\noindent provided that the boundary conditions on $g$ and $f_{\vc{Y}}$ are chosen to satisfy
\begin{equation}
\int\limits_{\partial\mathcal{Y}}\left(\mathbb{D}_{j}^{(1)}gf_{\vc{Y}}+\pd{g}{y_{j}}f_{\vc{Y}}\mathbb{D}_{ij}^{(2)}-
g\pd{f_{\vc{Y}}\mathbb{D}_{ij}^{(2)}}{y_{i}}\right)n_{j}\rd\vc{y}=0,
\label{eq:bcs}
\end{equation}
\noindent where $n_{j}$ is the $j^{\mathrm{th}}$ component of the
unit outward normal of $\partial\mathcal{Y}$. The forward equation
evolves a probability density forwards in time from a specified
initial density $f_{\vc{Y}}(\vc{y},0)$.

\subsection{Remarks}
\label{sec:Remarks}

\begin{enumerate}[wide=0pt,label=\emph{(\roman*)}]
\item For $f_{\vc{Y}}$ to be a probability density, its integral
  over $\mathcal{Y}$ must be equal to unity for all
  time, which places a constraint on $\mathbb{D}^{(0)}$. However, the continuity equation
  $\nabla\cdot\vc{U}_{t}\equiv 0$ was used to obtain
  $\mathbb{D}^{(0)}$, which implies that

  \begin{equation}
\int\limits_{\mathcal{Y}}\mathbb{D}^{(0)}f_{\vc{Y}}\rd\vc{y}=-\phi\,\int\limits_{\mathcal{Y}}\mathbb{E}_{\vc{Y}_{s}|\partial\Omega}[\vc{n}\cdot\vc{U}_{s}]\frac{f_{\vc{Y}|\partial\Omega}}{f_{\vc{Y}}}\rd\vc{y}=-\mathbb{E}[\vc{n}\cdot\vc{U}_{s}]=0,
\end{equation}

\noindent and means that the forcing term in \eqref{eq:kf} does
not affect the integral of $f_{\vc{Y}}$ by construction. 

\item Diffusive fluxes $-\alpha_{ij}\vc{n}\cdot\nabla Y_{t}^{j}$
  at the boundary contribute to drift in the evolution of the
  density $f_{\vc{Y}}$. A diffusive source (sink) of $Y_{t}^{i}$ at the boundary,
  conditional on a given value of $Y_{t}^{i}$, will transport $f_{\vc{Y}}$
  in the direction of positive (negative) $Y_{t}^{i}$. This
  process leads to drift, rather than the sources/sinks discussed
  in $(i)$ because diffusion irreversibly mixes boundary concentrations
  of $\vc{Y}_{t}$ with those that are already present inside the domain.
\item The familiar result that the variance $\|Y_{t}\|^{2}$ of a
  zero-mean scalar field ($n=1$) with diffusivity $\alpha$ on an
  insulated domain satisfies

  \begin{equation}
    \partial_{t}\|Y_{t}\|_{2}^{2}=-\alpha\|\nabla Y_{t}\|_{2}^{2},
    \label{eq:var}
  \end{equation}
  
  can be recovered by multiplying \eqref{eq:forward} (for $n=1$) by $y$
  and integrating with respect to $y\in (-\infty,+\infty)$.
  
\item $\mathbb{D}^{(0)}$, $\mathbb{D}^{(1)}$ and
  $\mathbb{D}^{(2)}$ in \eqref{eq:forward} involve unclosed terms
  and would therefore need to be modelled in order for
  \eqref{eq:forward} to be used prognostically (note that the
  surface density $f_{\vc{Y}|\partial\Omega}$ is not known a
  priori in general). Modelling in this regard must account for
  the effects of averaging over `degrees of freedom' that were not
  included in the state variable $\vc{Y}_{t}$. As with any
  coarse-grained representation, one's aim is to forecast a
  marginal distribution $f_{\vc{Y}}$ that is a good approximation
  to the corresponding projection of the full state of the system.
  
\item Expectations of the cross-gradient mixing terms
  $-\vc{\alpha}\nabla \vc{Y}_{t}\nabla \vc{Y}_{t}^{\top}$
  determine the diffusion coefficient $\mathbb{D}^{(2)}$ in
  \eqref{eq:D2}. The sign of $\mathbb{D}^{(2)}$ will be discussed
  in detail in \S\ref{sec:D2}. For many practical applications, it
  is reasonable expect $\mathbb{D}^{(2)}$ to be negative
  semidefinite, corresponding to the fact that down-gradient
  molecular transport homogenises $\vc{Y}_{t}$, which leads to
  greater certainty in the value of $\vc{Y}_{t}$. For example, in
  physical space, the eventual steady state of a scalar subjected
  to diffusion in an insulated domain will be uniform, as
  predicted by \eqref{eq:var}, which corresponds to a Dirac
  measure in the probability distribution.

\end{enumerate}

\section{The diffusion coefficient}
\label{sec:D2}

The properties of the diffusion coefficient $\mathbb{D}^{(2)}$
appearing in \eqref{eq:forward} are intriguing because they depend
on both the relative diffusivities $\vc{\alpha}$ of the observed
quantities $\vc{Y}_{t}$ and the correlation between gradients
$\nabla\vc{Y}_{t}$. It is therefore useful to understand when to
expect negative semidefinite $\mathbb{D}^{(2)}$ (which will be denoted
$\mathbb{D}^{(2)}\preceq 0$ or, equivalently,
$-\mathbb{D}^{(2)}\succeq 0$ when it is convenient to refer
to positive semidefinite matrices).

Begin by noting that outer products $\vc{v}\vc{v}^{\top}$ for
$\vc{v}\in\mathbb{R}^{n}$, are extreme rays and, therefore,
generators of the convex cone of positive semidefinite $n\times n$
matrices \citep{BleGboo2012a}. In particular, the matrix product
$\nabla\vc{Y}_{t}\nabla\vc{Y}_{t}^{\top}\succeq 0$, because it can
be represented as the sum of $d$ outer products (one for gradients
with respect to each spatial dimension
$X_{1},X_{2},\ldots,X_{d}$). Indeed, the expectation of such
products also produces positive semidefinite matrices, because
$\vc{u}^{\top}\mathbb{E}_{\vc{Y}_{t}}[\nabla\vc{Y}_{t}\nabla\vc{Y}_{t}^{\top}]\vc{u}=\mathbb{E}_{\vc{Y}_{t}}[(\vc{u}^{\top}\nabla\vc{Y}_{t})(\nabla\vc{Y}_{t}^{\top}\vc{u})]\geq
0$ for all nonzero $\vc{u}\in\mathbb{R}^{n}$ is a sum of squares,
which implies that
\begin{equation}
  \mathbb{E}_{\vc{Y}_{t}}[\nabla\vc{Y}_{t}\nabla\vc{Y}_{t}^{\top}]\succeq 0.
  \label{eq:sos}
\end{equation}

\begin{figure}[t]
    \begin{subfigure}[t]{0.4\textwidth}
        \centering
        \caption{}
        \scalebox{0.9}{\input{diag_D2}}
        \label{fig:D2a}
    \end{subfigure}
    \begin{subfigure}[t]{0.6\textwidth}
        \centering
        \caption{}
        \includegraphics[scale=1.0]{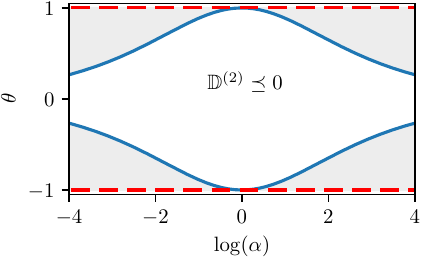}
        \label{fig:D2b}
    \end{subfigure}
    \caption{$(a)$ Violation of positive semidefiniteness in the
      generator $\vc{\alpha}\vc{v}\vc{v}^{\top}$ of
      $-\mathbb{D}^{(2)}$ when the quantities $\vc{Y}_{t}$ have
      different diffusivities $\vc{\alpha}$. $(b)$ The
      relationship between the ratio of diffusivities $\alpha$ and
      the correlation coefficient $\theta$ that ensures that
      $\mathbb{D}^{(2)}$ in \eqref{eq:D2_example} is negative
      semidefinite.}
\end{figure}

\noindent The (negative) diffusion coefficient $-\mathbb{D}^{(2)}$
in \eqref{eq:D2}, on the other hand, is effectively generated by
the outer product of the vectors $\vc{\alpha}\vc{v}$ and
$\vc{v}$. If the diffusivities of each observable quantity
$\vc{Y}_{t}$ are equal and nonnegative, such that
$\vc{\alpha}=\alpha\vc{I}_{n}$ for $\alpha\in\mathbb{R}_{\geq 0}$,
then $\vc{\alpha}\vc{v}$ and $\vc{v}$ point in the same direction
and $\mathbb{D}^{(2)}$ is negative semidefinite
($-\mathbb{D}^{(2)}\succeq 0$). More generally, however, when
$\vc{\alpha}\neq \alpha\vc{I}_{n}$, the outer product
$\vc{\alpha}\vc{v}\vc{v}^{\top}$ creates the possibility of
$\mathbb{D}^{(2)}$ being sign indefinite, as illustrated by the shaded region of figure
\ref{fig:D2a}. The determining factor in such cases are the
correlations between $\nabla\vc{Y}_{t}$, with respect to both the
sample space and the gradient directions, since it is possible for
the sum of sign indefinite matrices to be positive
semidefinite. Weaker correlations between the gradients, provide
stronger mitigation of the effects of unequal diffusivities in
$\vc{\alpha}$. For example, if $\vc{\alpha}$ is any nonnegative
diagonal matrix and the components of $\nabla\vc{Y}_{t}$ are
uncorrelated, such that
$\mathbb{E}_{Y}[\nabla\vc{Y}_{t}^{i}\cdot\nabla\vc{Y}_{t}^{j}]=0$
for $i\neq j$, then $\mathbb{D}^{(2)}\preceq 0$ is a negative
semidefinite diagonal matrix. More generally, it is also worth
noting that the scalar diffusion coefficients associated with the
marginal distributions of $Y_{t}^{1}, Y_{t}^{2},\ldots,Y_{t}^{n}$
correspond to the diagonal elements of $\vc{\alpha}$.

A simple two-dimensional example illustrates the combined effects
of correlation in the gradients $\nabla \vc{Y}_{t}$ and unequal
diffusivities in $\vc{\alpha}$. Let $\alpha_{11}=1$,
$\alpha_{22}=\alpha$ and
$\alpha_{12}=\alpha_{21}=0$ and assume, without loss of
generality, that
$\mathbb{E}_{\vc{Y}_{t}}[\nabla
Y_{t}^{1}]=\mathbb{E}_{\vc{Y}_{t}}[\nabla Y_{t}^{2}]=1$ to
normalise the problem. Define the correlation coefficient
$\theta:=\mathbb{E}_{\vc{Y}_{t}}[\nabla Y_{t}^{1}\cdot\nabla
Y_{t}^{2}]$, such that

\begin{equation}
  -\mathbb{D}^{(2)}=
  \begin{bmatrix}
    1 & \theta(1+\alpha)/2 \\
    \theta(1+\alpha)/2 & \alpha
  \end{bmatrix}.
  \label{eq:D2_example}
\end{equation}

\noindent from \eqref{eq:D2}. The (negative) symmetric diffusion coefficient
$-\mathbb{D}^{(2)}$ in \eqref{eq:D2_example} is positive semidefinite if its most negative eigenvalue
\begin{equation}
\lambda = \frac{ (1+\alpha) -\sqrt{(1+\alpha)^{2}+\theta^{2}(1+\alpha)^{2}-4\alpha} }{2},
\end{equation}
\noindent is nonnegative, which means that 
\begin{equation}
  |\theta|\leq \frac{2\sqrt{\alpha}}{1+\alpha},
  \label{eq:theta}
\end{equation}
\noindent is the required relation between correlation and
diffusivity that guarantees $-\mathbb{D}^{(2)}\succeq 0$ and is
illustrated in figure \ref{fig:D2b}. If $\theta\in \{-1,1\}$, any
difference in $\alpha$ from unity will lead to a sign indefinite
diffusion coefficient $\mathbb{D}^{(2)}$, as motivated in the text
below \eqref{eq:sos} and figure \ref{fig:D2a}. On the other hand,
for weak correlations $|\theta|<1$ the ratio of the diffusivities
$\alpha$ has to be either large or small to produce a sign
indefinite diffusion coefficient $\mathbb{D}^{(2)}$, as
illustrated by the grey regions in figure \ref{fig:D2b}. For
reference, the ratio of thermal diffusivity to mass diffusivity in
the oceans is around $100$ and can be regarded as corresponding to
$\alpha$ in the example above. Therefore, according to
\eqref{eq:theta}, in that case $|\theta|$ would need to be less
than approximately $0.2$ for $\mathbb{D}^{(2)}$ to be negative
semidefinite.

\begin{figure}
  \begin{center}
  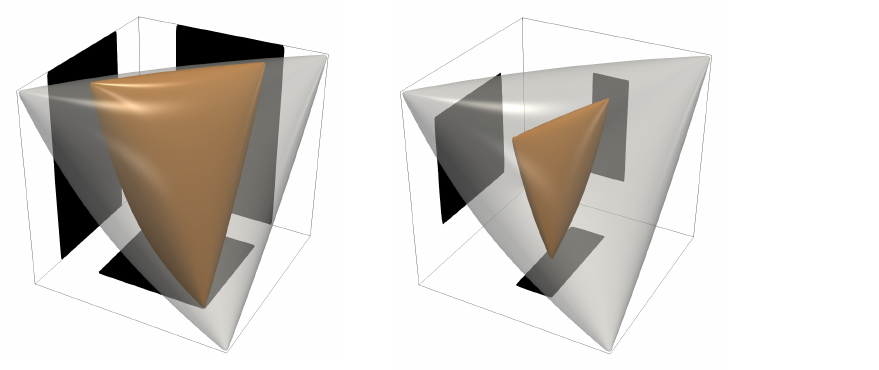
  \end{center}
  \caption{Spectrahedra corresponding to the condition that
    $-\mathbb{D}^{(2)}\succeq 0$ for correlation coefficients
    $\theta_{1}$, $\theta_{2}$ and $\theta_{3}$ and different
    relative diffusivities $\alpha_{2}$ and $\alpha_{3}$ in
    \eqref{eq:D2_example3d}. In $(a)$ $\alpha_{2}=1$ and
    $\alpha_{3}=10$ and in $(b)$ $\alpha_{2}=0.1$ and
    $\alpha_{3}=10$. The red regions correspond to
    $\mathbb{D}^{(2)}\preceq 0$, whilst the larger grey region
    corresponds to the feasibility requirement \eqref{eq:sos} for
    the correlations. The range of all axes is $[-1,1]$.}
\label{fig:spectrahedra}
\end{figure}

When $\vc{\alpha}$ is prescribed, it is useful to know the
correlation coefficients $\vc{\theta}\in\mathbb{R}^{n(n-1)/2}$
that produce negative semidefinite diffusion coefficients
$\mathbb{D}^{(2)}$. In such cases, the condition that
$-\mathbb{D}^{(2)}\succeq 0$ can be represented by spectrahedra,
which are formally defined as the intersection of the convex cone
of positive semidefinite matrices \eqref{eq:sos} in $\mathbb{R}^{n\times n}$ with
an affine subspace \citep{BleGboo2012a}. In three dimensions, setting
$\alpha_{11}=1$, $\alpha_{22}=\alpha_{2}$ and $\alpha_{33}=\alpha_{3}$,

\begin{equation}
-\mathbb{D}^{(2)}=    
  \begin{bmatrix}
    1 & \theta_{3}(1+\alpha_{2})/2 & \theta_{2}(1+\alpha_{3})/2 \\
    \theta_{3}(1+\alpha_{2})/2 & \alpha_{2} & \theta_{1}(\alpha_{2}+\alpha_{3})/2 \\
    \theta_{2}(1+\alpha_{3})/2 & \theta_{1}(\alpha_{2}+\alpha_{3})/2 & \alpha_{3}
  \end{bmatrix},
  \label{eq:D2_example3d}
\end{equation}

\noindent where $\theta_{1}$, $\theta_{2}$ and $\theta_{3}$ are
correlation coefficients. Spectrahedra corresponding to
$-\mathbb{D}^{(2)}\succeq 0$ are shown in red in figure
\ref{fig:spectrahedra}. The larger spectrahedron shown in grey
corresponds to \eqref{eq:sos}, which constrains the possible
values of $\vc{\theta}$. Values of $\vc{\theta}$ that lie between
the red and the grey regions therefore correspond to permissible
but sign indefinite diffusion coefficients $\mathbb{D}^{(2)}$. In
figure \ref{fig:spectrahedra}a, $\alpha_{2}=1$ and
$\alpha_{3}=10$, which leads to a narrower range of $\theta_{1}$
and $\theta_{2}$ values for which $-\mathbb{D}^{(2)}\succeq 0$. In
figure \ref{fig:spectrahedra}b $\alpha_{2}=0.1$ and $\alpha_{3}=10$ which
narrows the range further, and increases the possibility that
$\mathbb{D}^{(2)}$ is sign indefinite.

\section{Example applications}
\label{sec:examples}

The following subsections illustrate aspects of the results derived
in \S\ref{sec:gov} by providing example applications.

\subsection{Advection and diffusion by the ABC flow}
\label{sec:ABC}

An Arnold-Beltrami-Childress (ABC) flow is the three dimensional
divergence-free velocity field
\begin{equation}
  \boldsymbol{U} :=
    (\sin(X_3) + \cos(X_2),\ 
    \sin(X_1) + \cos(X_3),\
    \sin(X_2) + \cos(X_1))^{\top}.
\end{equation}
\noindent It is an exact solution of the Euler equations in a
periodic domain and is known to exhibit chaotic streamlines
\citep{DomTjfm1986a}, which makes it an ideal candidate to study
mixing. Couched in the format of \eqref{eq:gov}, the evolution of
a passive scalar concentration $Y_t$ due to combined effects of
advection by $\vc{U}$ and diffusion is

\begin{equation}
\od{Y_t}{t} = -\boldsymbol{U} \cdot \nabla Y_t + \alpha \Delta Y_t,
\label{eq:Scalar_Advection_ODE}
\end{equation}

\noindent where $\alpha$ denotes the constant scalar diffusivity. The
corresponding forward Kolmogorov equation governing the evolution
of $f_{Y}(y,t)$ is given by \eqref{eq:forward}:

\begin{equation}
\frac{\partial }{ \partial t} f_{Y} = -\alpha\frac{\partial^2 }{ \partial y^2} \big( \underbrace{\mathbb{E}_{Y_t}[|\nabla Y_t|^2]}_{=:\mathbb{D}^{(2)}} f_{Y} \big).
\label{eq:fK_ABC_flow}
\end{equation}

The coefficients $\mathbb{D}^{(0)}$ and
$\mathbb{D}^{(1)}$ from \S\ref{sec:backward} are both zero
because the periodic domain does not have boundaries. Although
\eqref{eq:fK_ABC_flow} cannot be solved, as $\mathbb{D}^{(2)}$ is
unknown, we can examine estimates of terms by
solving \eqref{eq:Scalar_Advection_ODE} numerically. Choosing an initial condition
consisting of a front separating two regions of different
concentration $Y_{0} = \tanh(10 \vc{X})$ and setting $\alpha=5$, we numerically integrate
with respect to $t\in[0,4]$. Then, by approximating the density
$f_{Y}$ with a histogram (as detailed in
\cite{Supporting_Examples}) we estimate the terms in \eqref{eq:fK_ABC_flow}.

\begin{figure}[t]
    \centering
    \begin{subfigure}[t]{0.24\textwidth}
        \centering
        \caption{$t=0.5$}
        \includegraphics[scale=0.3]{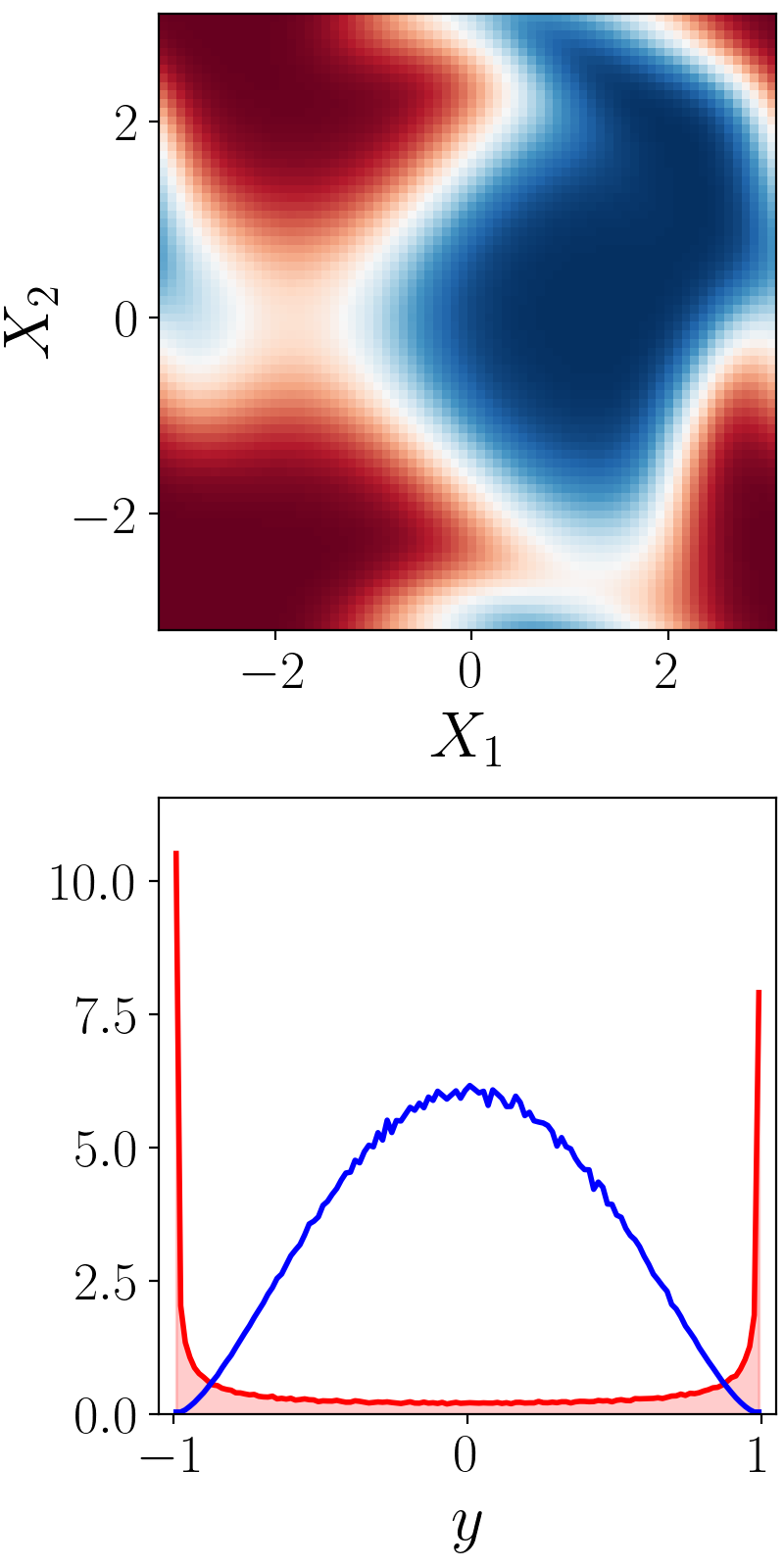}
    \end{subfigure}
    \begin{subfigure}[t]{0.24\textwidth}
        \centering
        \caption{$t=1$}
        \includegraphics[scale=0.3]{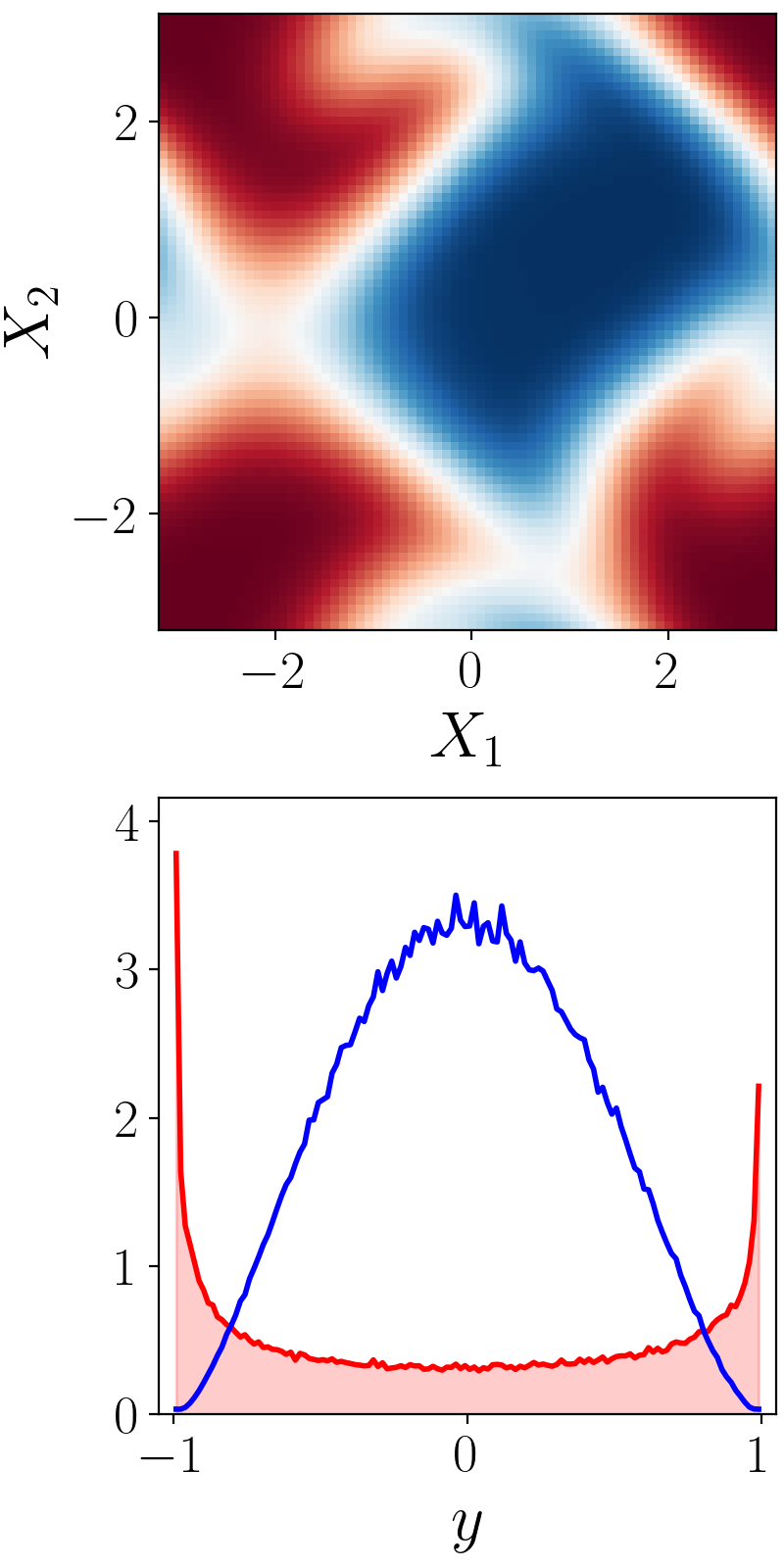}
    \end{subfigure}
    \begin{subfigure}[t]{0.24\textwidth}
        \centering
        \caption{$t=2$}
        \includegraphics[scale=0.3]{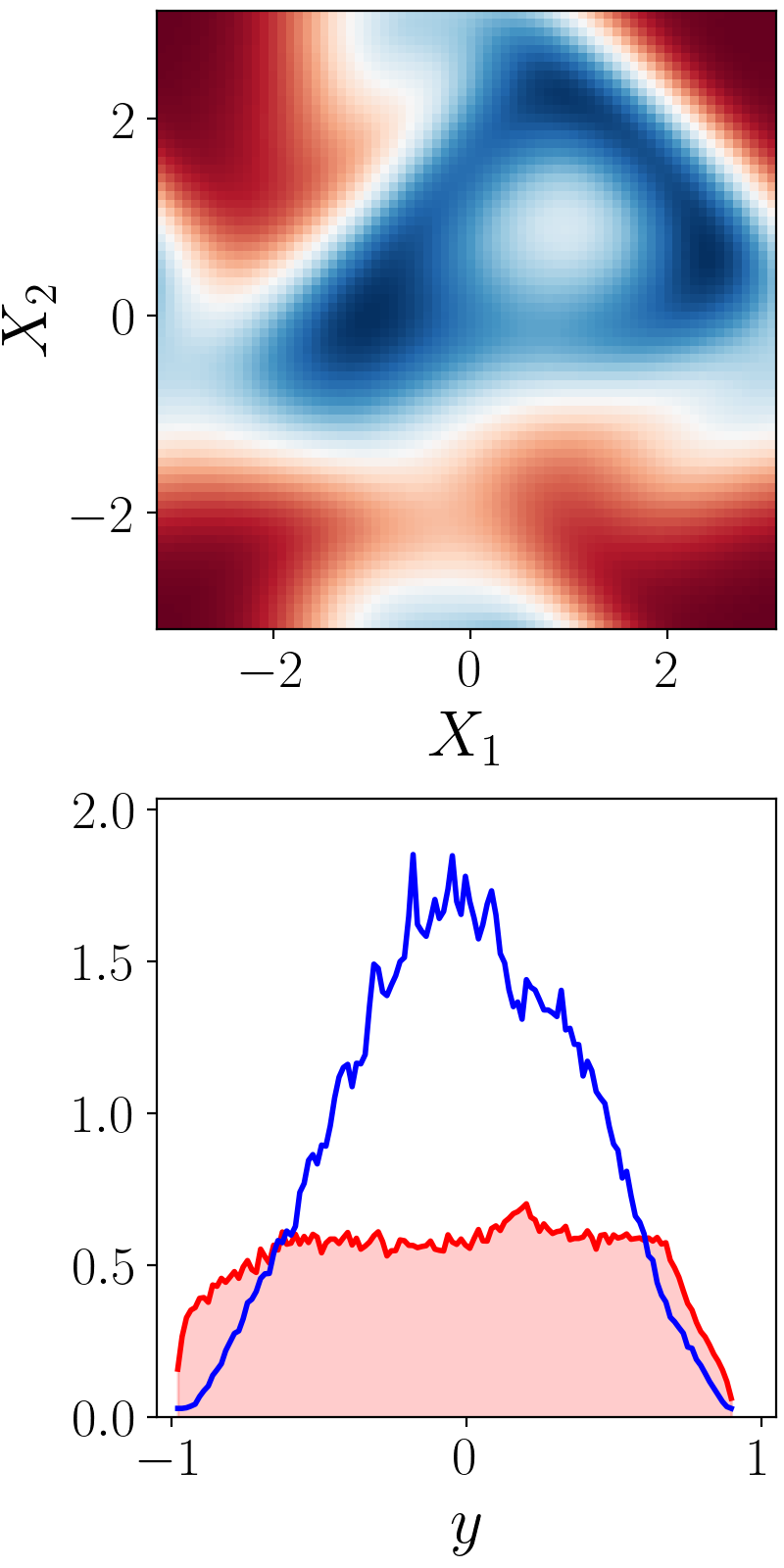}
    \end{subfigure}
    \begin{subfigure}[t]{0.24\textwidth}
        \centering
        \caption{$t=4$}
        \includegraphics[scale=0.3]{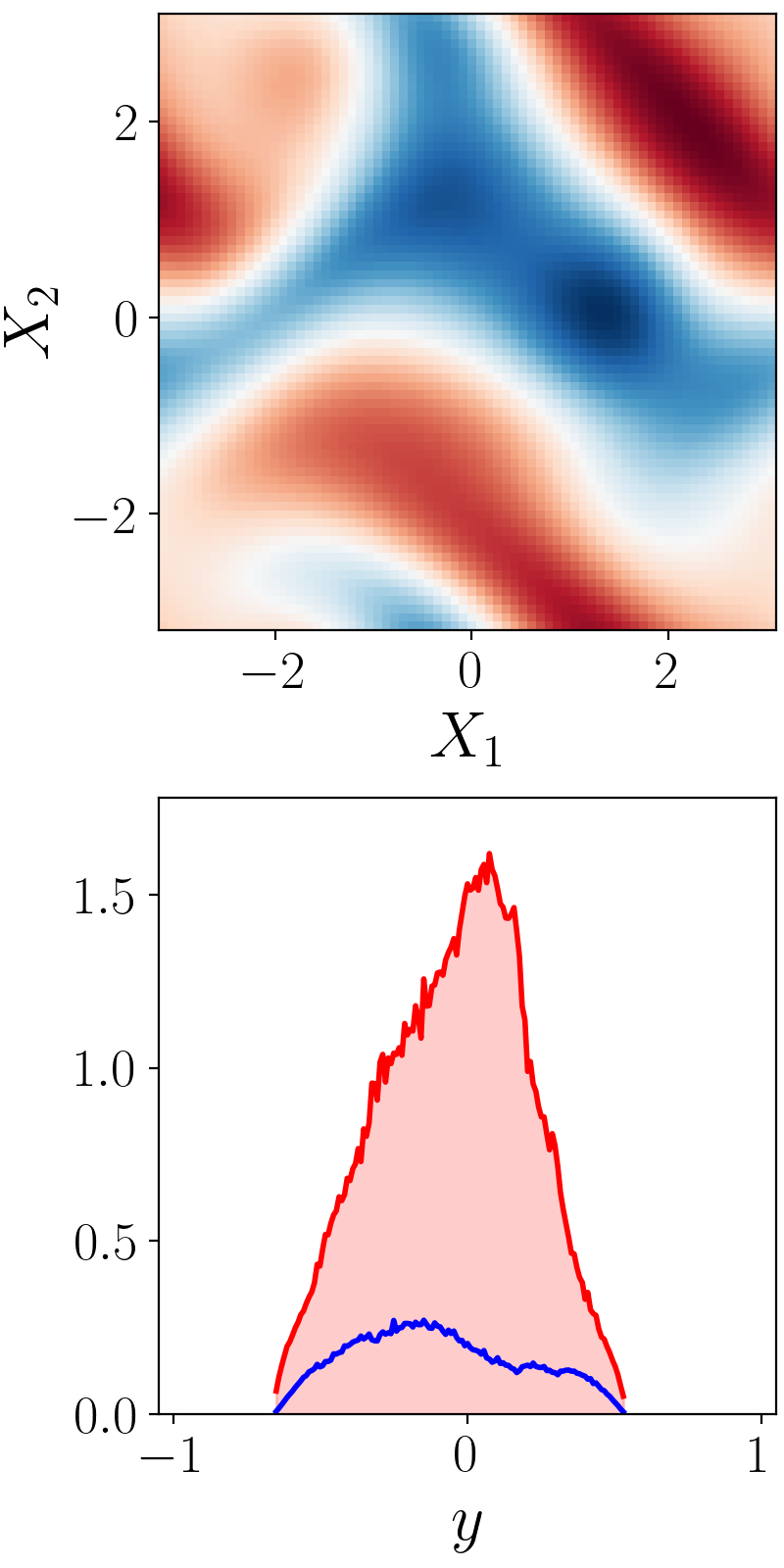}
    \end{subfigure}
    \caption{(top) Time evolution of a cross section of the concentration field $Y_{t}$ and (bottom) its corresponding global density $f_{Y}(y,t)$ (shaded) and (negative) diffusion coefficient $\mathbb{E}_{Y_t}[|\nabla Y_t|^2]$.}
    \label{fig:ABC_flow}
\end{figure}

A time evolution of a cross section of the scalar field $Y$
alongside its corresponding density $f_{Y}(y,t)$ and the
conditional expectation $\mathbb{E}_{Y_t}[|\nabla Y_t|^2]$ are
shown in figure \ref{fig:ABC_flow}. At $t=0.5$ the scalar field,
consisting of two regions of approximately uniform concentration
separated by a relatively sharp interface, is represented in
probability space by spikes at $y = \pm 1$. As the concentration
field is mixed irreversibly, the amplitude of these spikes reduces
and the density associated with values of $y$ in the vicinity of
zero increases. Mixing subsequently homogenises the scalar field
to the extent that no evidence of the initial distribution remains
by $t=2$. As there is no injection of concentration to balance the
homogenising effects of diffusion, $Y_t$ will ultimately tend to a
constant uniformly over the domain and the density $f_{Y}(y,t)$ in
the limit $t\rightarrow\infty$ would tend to a Dirac distribution.

The expectation $\mathbb{E}_{Y_t}[|\nabla Y_t|^2]$, which
quantifies the amount of mixing and, therefore, negative diffusion
in \eqref{eq:fK_ABC_flow} is maximised by $y=0$. For this example,
scalar concentrations that are less probable are therefore
associated with the most mixing. This observation is not
surprising because, as illustrated in figure \ref{fig:example1d},
large gradients are associated with small probability
densities. In particular, $\mathbb{E}_{Y_t}[|\nabla Y_t|^2]$ is
zero for the minimum ($y=-1$) and maximum ($y=1$) values of
$Y_{t}$, which necessarily correspond to local extrema, for which
$\nabla Y_t=\vc{0}$. In the absence of diffusion at these values
of $y$, the density $f_{Y}$ consequently maintains compact support
(i.e. mixing interpolates and cannot produce values that lie
outside the range of values that were there in first place).

\subsection{Stochastic boundary conditions}
\label{sec:1D_Diffusion}

In order to sustain a finite variance of a diffusive scalar field
it necessary to apply a forcing. Given that a wide variety of
physical systems are sustained by a boundary forcing, often
characterised by substantial fluctuations and uncertainty, it is
natural to consider how these forces manifests in the evolution
equation governing the system's probability density. A simple
example of such a problem that is sufficient to highlight several
of the terms discussed in \S\ref{sec:gov} is the diffusion of heat
through a one dimensional rod of unit length forced by
Ornstein-Uhlenbeck processes (i.e. normally distributed thermal
fluctuations) at the boundary. When cast in the form of
\eqref{eq:gov}, for unit thermal diffusivity ($\alpha=1$) the
temperature $Y_{t}$ evolves according to
\begin{equation}
    \od{Y_t}{t} = \partial^{2}_{X}Y_t,\quad \mathrm{for}\quad X\in\mathcal{X}:=[0,1],
    \label{eq:1D_diffusion}
\end{equation}
where $\partial^{2}_{X}Y_{t}$ is to be regarded as an unknown
`random' variable and values of $Y_{t}$ for points on the boundary
$\partial\Omega$ (corresponding to $\partial\mathcal{X}=\{0,1\}$)
evolve according to
\begin{equation}
  \rd Y_t(\omega) = -aY_t(\omega)\rd t + \sigma \rd W_t(\omega),\quad \mathrm{for}\quad \omega\in\partial\Omega,
  \label{eq:oup}
\end{equation}
where $a,\sigma$ are real constants and $W_t(\omega)$ is a Wiener
process.

\begin{figure}[t]
    \centering
    \begin{subfigure}[t]{0.71\textwidth}
      \caption{}
       \includegraphics[scale=0.3]{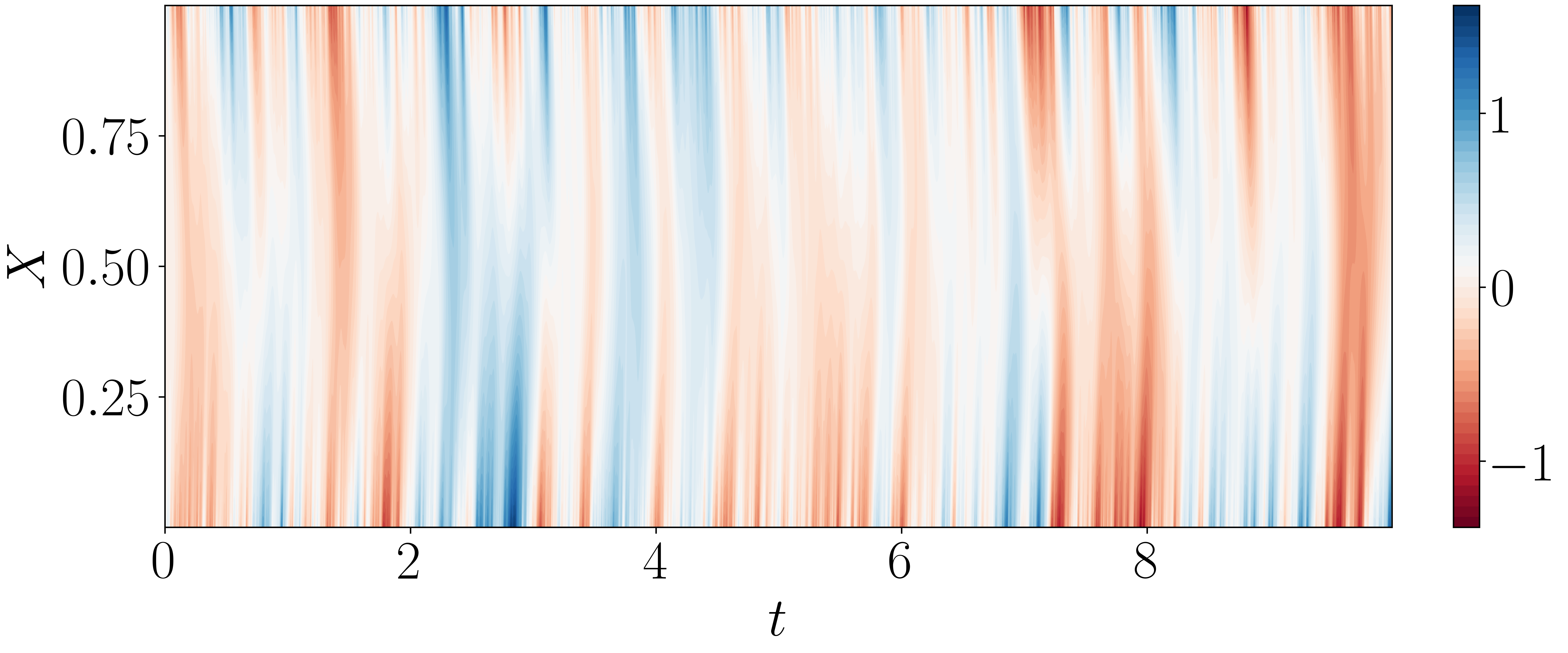}
    \label{fig:Diffusion_1D_Space_Time}
    \end{subfigure}
    \begin{subfigure}[t]{0.28\textwidth}
      \caption{}
       \includegraphics[scale=0.3]{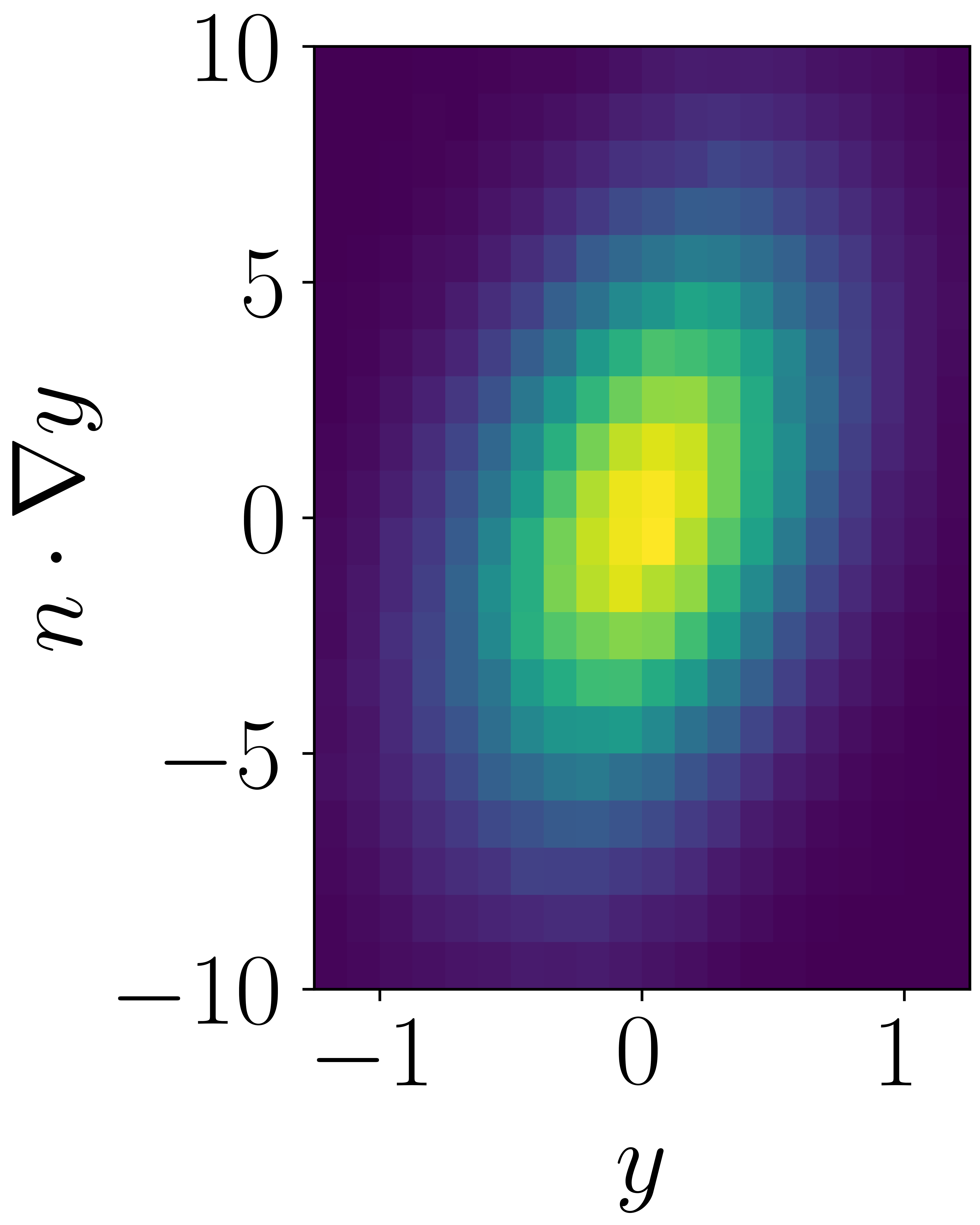}
    \label{fig:Diffusion_1D_Joint_Density}
    \end{subfigure}
    \caption{$(a)$ Space-time plot of $Y_t$ for the one diffusion equation \eqref{eq:1D_diffusion} forced by an Ornstein-Ulhenbeck process \eqref{eq:oup} with $a=5,\sigma=1$, and $(b)$ the conditional joint density $f_{Y, \partial_{n}Y|\partial\Omega}$. The positive covariance of this plot reflects the tendency of heat to flow down gradients at the boundaries.}
    \label{fig:Diffusion_1D}
\end{figure}

Figure \ref{fig:Diffusion_1D_Space_Time} shows a space-time plot
of the system. At the boundaries the temperature fluctuates on the
integral time-scale ($1/a$) of the Ornstein-Ulhenbeck processes
\eqref{eq:oup}.  Towards the centre of the domain the effects of
diffusion reduce the amplitude of the fluctuations.

The forward Kolmogorov equation, which describes how the
probability density $f_{Y}(y,t)$ evolves in time is given by
\eqref{eq:kf}:

\begin{equation}
    \frac{\partial }{ \partial t} f_Y = -\frac{\partial}{ \partial y} \big( \underbrace{ 2\mathbb{E}_{Y_t}[ \partial_{n}Y_t ] \frac{f_{Y|\partial\Omega}}{f_{Y}}}_{=:\mathbb{D}^{(1)}} f_{Y} \big) - \frac{\partial^2 }{ \partial y^2} \big( \underbrace{\mathbb{E}_{Y_t}[|\partial_{X} Y_t|^2]}_{=:\mathbb{D}^{(2)}} f_{Y} \big),
    \label{eq:fK_diffusion}
\end{equation}
where $f_{Y|\partial\Omega}$ corresponds to the density of $Y_t$
when it is sampled from the boundary. According to \eqref{eq:oup}
$Y_{t}$ sampled at the boundary is an Ornstein-Uhlenbeck process
and therefore has a stationary density given by
$f_{Y|\partial\Omega}(y)=\sqrt{a/(2\pi\sigma^{2})}\,\mathrm{exp}(-ay^{2}/(2\sigma^{2}))$
\citep{GarCboo1989a}. The factor $\phi=2$ in the first term on the
right hand side results from the fact that domain's boundary
consists of two points (cf. figure \ref{fig:ratios}).

As there is no advection in \eqref{eq:1D_diffusion} the
coefficient $\mathbb{D}^{(0)}$ is zero. However, due to conduction
and, in turn, boundary fluxes of heat, \eqref{eq:fK_diffusion} has
non-zero drift and diffusion terms. Although
\eqref{eq:fK_diffusion} cannot be solved prognostically without a
model for the unknown coefficients, we can estimate the
coefficients numerically by constructing an ensemble satisfying
\eqref{eq:1D_diffusion} and \eqref{eq:oup} as detailed in the
supporting example code \cite{Supporting_Examples}.

\begin{figure}[t]
    \begin{center}
    \includegraphics[scale=0.425]{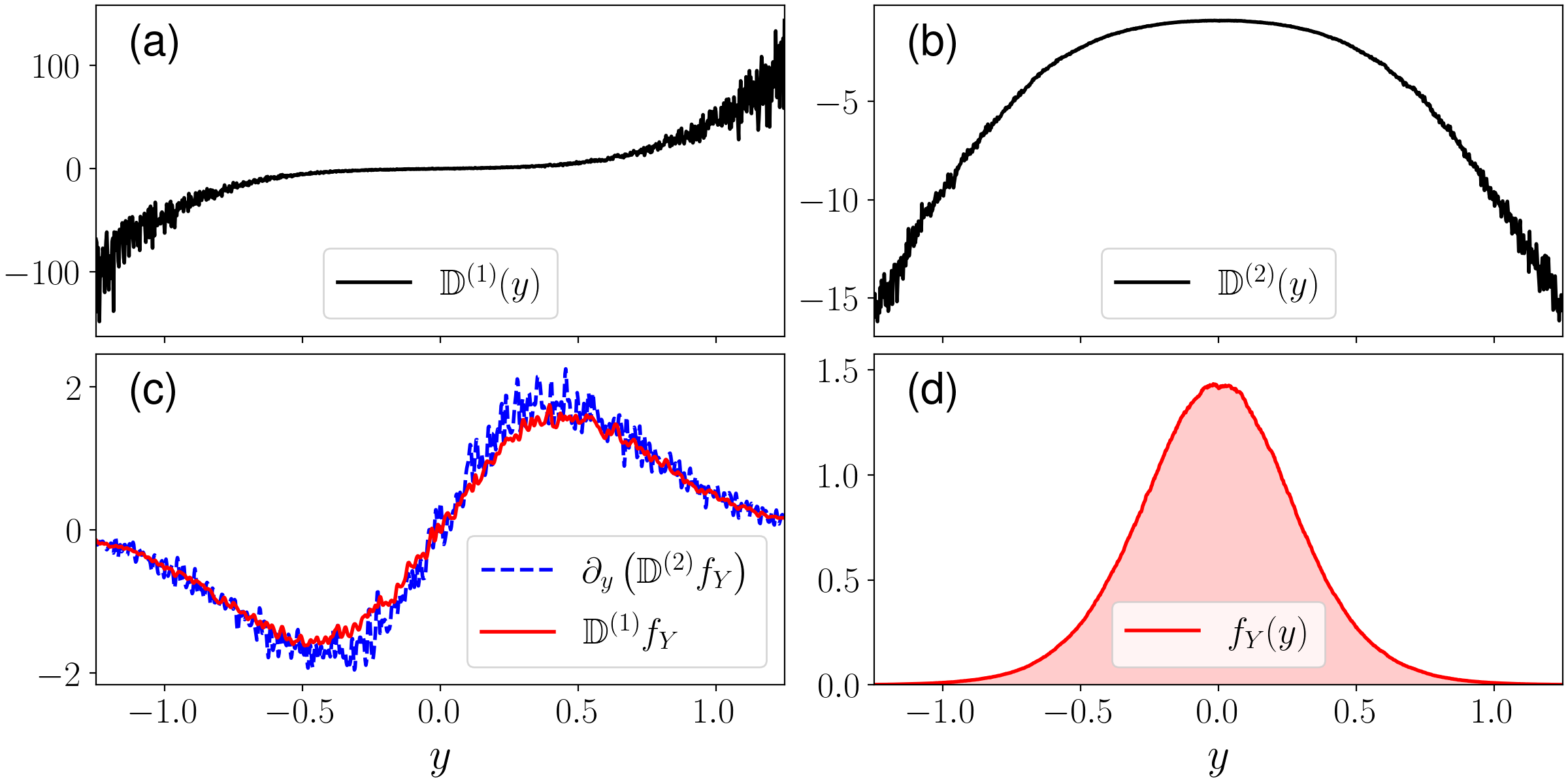}
     \end{center}
     \caption{$(a,b,d)$ The drift $\mathbb{D}^{(1)}$, diffusion
       $\mathbb{D}^{(2)}$ and probability density $f_Y(y)$,
       respectively, of the forward Kolmogorov equation
       corresponding to a stationary state of the system
       \eqref{eq:1D_diffusion} with $a=5,\sigma=1$. (c) A
       numerical verification of the balance between boundary
       forcing and mixing.}
    \label{fig:Diffusion_1D_Coefficients}
\end{figure}

Figure \ref{fig:Diffusion_1D_Coefficients} shows the drift
$\mathbb{D}^{(1)}$ and diffusion $\mathbb{D}^{(2)}$ coefficients,
as well as the stationary density $f_{Y}(y)$, for which use has
been made of time averaging. In \S\ref{sec:ABC}, molecular
diffusion acted to homogenise the scalar field $Y_t$ in the ABC
flow and drive its corresponding density towards a Dirac
measure. Were the value of $Y_t$ at the boundaries in this example
fixed at zero, the distribution would, due to the negative
diffusion $\mathbb{D}^{(2)}$ (figure
\ref{fig:Diffusion_1D_Coefficients}d), also tend towards a Dirac
measure at zero irrespective of the precise initial
conditions. Instead, random forcing at the system's boundary
creates variance that balances the destruction of variance
described by $\mathbb{D}^{(2)}$. To illustrate how, figure
\ref{fig:Diffusion_1D_Space_Time} shows the conditional joint
density $f_{Y, \partial_{n} Y_{t}|\partial\Omega}$ of $Y_t$ and
its boundary-normal gradient $\partial_{n}Y_{t}$. The conditional
density reveals an expected positive correlation between $Y_t$ and
$\partial_{n}Y_t$, which means that heat flux into the domain is
typically accompanied by positive temperatures at the
boundary. Therefore $\mathbb{D}^{(1)}\lessgtr 0$ for
$y\lessgtr 0$, as shown in figure
\ref{fig:Diffusion_1D_Coefficients}a, which implies that
$\mathbb{D}^{(1)}$ corresponds to divergent transport of
probability density away from the origin.

Computing the balance of the terms in the right-hand side of
\eqref{eq:fK_diffusion}, as shown in figure
\ref{fig:Diffusion_1D_Coefficients}, verifies
\eqref{eq:fK_diffusion}. It is worth noting that while
incorporating uncertainty into \eqref{eq:1D_diffusion} required a
Monte-Carlo approach to be employed in order to estimate $f_Y(y)$,
such boundary conditions pose no particular additional difficulty
in \eqref{eq:fK_diffusion}, which accommodates stochastic forcing
naturally.

\subsection{The Boussinesq equations}
\label{sec:bou}

The following example demonstrates that coordinates, regarded here
as functions $\vc{X}$ acting on the domain $\Omega$, can be
included in the state vector $\vc{Y}_{t}$. For instance, if one
wishes to understand probability distributions over horizontal
slices of a domain, the vertical coordinate can be included in
$\vc{Y}_{t}$. More generally, other functions, such as
geopotential height, could be incorporated into $\vc{Y}_{t}$. The
equations developed in \S\ref{sec:gov} are therefore very general
in being able to generate equations that are specific to a
particular problem and circumvent the need for case-specific
derivations.

In buoyancy-driven flows the vertical coordinate (or, more
generally, geopotential height) plays an important role in
corresponding to the direction of gravity. Therefore, let
$\vc{Y}_{t}:=(B_{t},W_{t},Z_{t})^{\top}$, where $B_{t}$ is the
buoyancy, $W_{t}$ is the vertical velocity and $Z_{t}=X_{3}$ is
the (time independent) vertical coordinate. In a Boussinesq
context, the pointwise deterministic equations governing the
behaviour of $B_{t}$ and $W_{t}$ are
\begin{equation}
    \od{B_{t}}{t}=-\vc{U}_{t}\cdot\nabla {B}_{t} -\alpha_{1}\Delta B_{t},\quad
    \od{W_{t}}{t}=B_{t}-\partial_{z}P_{t} -\vc{U}_{t}\cdot\nabla {W}_{t}+\alpha_{2}\Delta W_{t},
\end{equation}
\noindent where $\vc{U}_{t}$ is a solenoidal velocity field that
is assumed to vanish at the boundary, which means that
$\mathbb{D}^{(0)}\equiv 0$. Writing
$\rd Z_{t}/\rd t=W_{t}-\vc{U}_{t}\cdot\nabla Z_{t}=0$ and noting that
$\alpha_{11}=\alpha_{1}$, $\alpha_{22}=\alpha_{2}$, $\alpha_{33}=0$ and
$\alpha_{ij}=0$ for $i\neq j$ implies that 
\begin{equation}
  \mathbb{D}^{(1)}=
    \begin{pmatrix}
    0\\
    b\\
    w
  \end{pmatrix}
-  
  \mathbb{E}_{\vc{Y}_{t}}
    \begin{bmatrix}
    0\\
    \partial_{Z} P_{t} \\
    0
  \end{bmatrix}
  +
  \phi\,\mathbb{E}_{\vc{Y}_{t}|\partial\Omega}
\begin{bmatrix}
  \alpha_{1}\,\vc{n}\cdot\nabla B_{t}\\
  \alpha_{2}\,\vc{n}\cdot\nabla W_{t}\\
  0
\end{bmatrix} \frac{f_{\vc{Y}|\partial\Omega}}{f_{\vc{Y}}},
\label{eq:bou_D1}
\end{equation}
\noindent and 
\begin{equation}
  \mathbb{D}^{(2)}=-\frac{1}{2}
  \mathbb{E}_{\vc{Y}_{t}}
    \begin{bmatrix}
    2\alpha_{1}|\nabla B_{t}|^{2} & (\alpha_{1}+\alpha_{2})\nabla B_{t}\cdot\nabla W_{t} & \alpha_{1} \partial_{Z}B_{t}\\
    (\alpha_{1}+\alpha_{2})\nabla B_{t}\cdot\nabla W_{t} & 2\alpha_{2} |\nabla W_{t}|^{2} & \alpha_{2} \partial_{Z}W_{t}\\
    \alpha_{1} \partial_{Z}B_{t} & \alpha_{2} \partial_{Z}W_{t} & 0 \\
  \end{bmatrix}.
\label{eq:bou_D2}
\end{equation}
\noindent In $\mathbb{D}^{(1)}$, $b$ is responsible for drift in
the $w$ direction, because buoyancy increases $W_{t}$ and, in
turn, $w$ is responsible for drift in the $z$ direction, because
vertical velocity increases $Z_{t}$. When interpreting the latter,
it should be borne in mind that for the closed domain in this
example $\mathbb{E}_{Z}[W_{t}]=0$, which, therefore, does not
affect the marginal distribution of $Z_{t}$ (i.e. the domain does
not change its shape over time). The unknown
conditionally-averaged vertical pressure gradient affects the
evolution of the joint density in the same way as
$b$. Additionally, the remaining terms in $\mathbb{D}^{(1)}$
account for boundary fluxes of $B_{t}$ and $W_{t}$, scaled by the
relative size of the boundary.

While the drift $\mathbb{D}^{(1)}$ is responsible for moving and
stretching the joint density, the diffusion coefficient
$\mathbb{D}^{(2)}$ accounts for the effects of irreversible
mixing. As discussed in \S\ref{sec:D2}, $\mathbb{D}^{(2)}$ is
expected to be negative semidefinite ($\mathbb{D}^{(2)}\preceq 0$)
in most applications, which means that it typically represents
antidiffusion. In particular, if $\alpha_{1}=\alpha_{2}=1$ then
$\mathbb{D}^{(2)}\preceq 0$ can be guaranteed. For other
combinations of $\alpha_{1}$ and $\alpha_{2}$ whether
$\mathbb{D}^{(2)}\preceq 0$ depends on the correlations between
$\nabla B_{t}$, $\nabla W_{t}$ and $\nabla Z$. The gradients,
$\nabla B_{t}$ and $\nabla W_{t}$ are a priori unknown and
therefore require closure, which would involve postulating their
dependence on the joint density $f_{\vc{Y}}$ and/or the
independent variables $\vc{y}:=(b, w, z)^{\top}$.

\begin{figure}
  \begin{center}
  \input{diag_leibniz}
  \end{center}
  \caption{Example marginal density $f_{Z}$, based on volume, and
    conditional marginal density $f_{Z|\partial\Omega}$, based on
    bounding surface area, for a cylinder of unit radius and
    length, capped with two hemispheres. The conditional density
    $f_{Z|\partial\Omega}(z)$ is proportional to the rate at which
    the bounding surface area below $z$ increases. It is uniform
    because the reduction in circumference in the spherical caps
    is balanced by an increase in $\gamma$, where
    $\mathrm{tan}(\gamma)$ corresponds to the relative amount of
    surface that is perpendicular to the vertical direction. For
    the geometry shown in this example, the total volume and
    surface area are $\mu^{3}(\mathcal{X})=\frac{7}{3}\pi$ and
    $\mu^{2}(\partial\mathcal{X})=6\pi$, respectively.}
\label{fig:leibniz}
\end{figure}
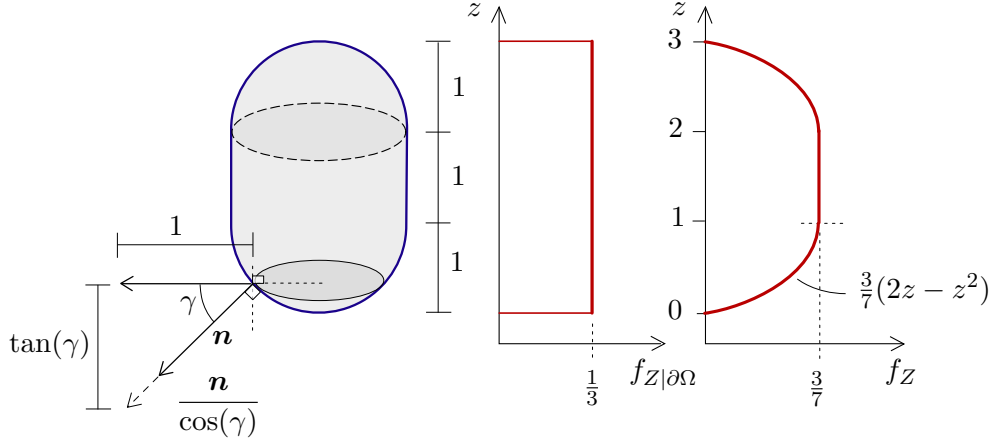

This example demonstrates that the framework developed in
\S\ref{sec:gov} for control volumes can be used immediately to
generate more specific and finer-grained equations over space. On
the other hand, integration of local equations for probability
density over submanifolds of an arbitrary domain would need to be
performed carefully in order to correctly account for fluxes
through the boundary \citep[see][for the analogous challenge in
analysing turbulent entrainment]{ReeMjfm2021a}. To see that
\eqref{eq:bou_D1} and \eqref{eq:bou_D2} are correct in this
regard, obtain an equation for $\mathbb{E}_{Z}[B]$ (i.e. the
average buoyancy at each height) by multiplying the forward
Kolmogorov equation by $b$ and integrating with respect to $b$ and
$w$:
\begin{equation}
  \pd{\mathbb{E}_{Z}[B_{t}]}{t}f_{Z} =\phi\alpha_{1}\mathbb{E}_{Z|\partial\Omega}[\vc{n}\cdot\nabla B_{t}]f_{Z|\partial\Omega}+\pd{}{z}\left(\mathbb{E}_{Z}[\alpha_{1}\partial_{Z}B_{t}-W_{t}B_{t}]f_{Z}\right).
\label{eq:dBdz}
  \end{equation}
  \noindent Now recall the definition of $\phi$ from
  \eqref{eq:phi} and note that $\mu^{d}(\mathcal{X})f_{Z}(z)$ is
  the rate at which the domain's volume beneath the height $z$
  increases as a function of $z$, which corresponds to the area
  associated with the horizontal slice
  $\mathcal{X}_{Z}(z)$. Similarly,
  $\mu^{d-1}(\partial\mathcal{X})f_{Z|\partial\Omega}(z)$ is the
  rate at which the domain's surface area beneath the height $z$
  increases as a function of $z$. The rate at which the surface
  area increases is proportional to the length of the boundary at
  the given height $z$, but also accounts for the angle
  $\gamma\in(-\pi/2,\pi/2)$ between $\vc{n}$ and the horizontal
  plane, as illustrated in figure \ref{fig:leibniz}. Parts of a
  bounding surface that are persistently perpendicular to the
  vertical direction, such that $\gamma=\pm\pi/2$ correspond to a
  Dirac measure in $f_{Z|\partial\Omega}$ and would therefore need
  to be handled separately or as a generalised function, because
  they cannot be represented as a density in the conventional
  sense.
  
  Using the information above, \eqref{eq:dBdz} can be expressed as
\begin{equation}
 \int\limits_{\mathcal{X}_{Z}(z)}\pd{B_{t}}{t}\rd\vc{x}=\int\limits_{\partial\mathcal{X}_{Z}(z)}\frac{\vc{n}\cdot\nabla B_{t}}{\mathrm{cos}(\gamma)}\rd\vc{x}+
  \frac{\partial}{\partial z}\int\limits_{\mathcal{X}_{Z}(z)}\alpha_{1}\pd{B_{t}}{z}-W_{t}B_{t}\rd\vc{x}.
\label{eq:b_average}
\end{equation}

\noindent The second and third terms in \eqref{eq:b_average} are
produced by the Leibniz integral rule for commuting the divergence
$\nabla\cdot$ and integration over $\mathcal{X}_{Z}(z)$. In
particular, the horizontal components of
$\vc{n}/\mathrm{cos}(\gamma)$ comprise the outward unit normal to
$\partial\mathcal{X}_{Z}(z)$ in the horizontal plane, with respect
to which the remaining component accounts for the relative size of
the vertical component of $\vc{n}$ \citep{ReeMjfm2021a}. (Un)applying
the divergence theorem to the diffusive fluxes over the horizontal
slice $\mathcal{X}_{Z}(z)$ and using the Leibniz integral rule to
commute $\partial_{z}$ with integration over $\mathcal{X}_{Z}(z)$
therefore yields
\begin{equation}
 \int\limits_{\mathcal{X}_{Z}(z)}\pd{B_{t}}{t}\rd\vc{x}=\int\limits_{\mathcal{X}_{Z}(z)}\alpha_{1}\Delta B_{t}-\pd{}{z}(W_{t}B_{t})\rd\vc{x},
\label{eq:leibniz}
\end{equation}
which corresponds to integration of the advection-diffusion
equation governing $B_{t}$ over a horizontal slice
$\mathcal{X}_{Z}(z)$ with zero velocity at the bounding surface.

\subsection{Available potential energy}

Available potential energy is the part of potential energy of a
body of fluid that is theoretically `available' for conversion
into kinetic energy. The other part, known as background potential
energy, cannot be converted into kinetic energy and is typically
associated with a stable equilibrium state
\cite{LorEtel1955a}. For example, kinetic energy cannot typically
be extracted from a stably stratified environment (in which a
fluid's density decreases with height) and therefore possesses
zero available potential energy.

When divided by the volume of the domain, the potential energy of
the fluid modelled in \S\ref{sec:bou} corresponds to the expectation of the
product $-Z_{t}B_{t}$ \citep[see, for example][]{WinKjfm1995a}:
\begin{equation}
  -\mathbb{E}\left[B_{t}Z_{t}\right]=-\int\limits_{\mathcal{Y}}bz f_{\vc{Y}}(\vc{y},t)\rd \vc{y},
  \label{eq:bz}
\end{equation}
\noindent where $f_{\vc{Y}}(\vc{y})$ is the joint density for the
variables $\vc{Y}_{t}:=(B_{t},W_{t},Z_{t})^{\top}$ evaluated at $\vc{y}:=(b,w,z)^{\top}$ and the minus
sign accounts for relatively warm parcels of fluid having greater
potential energy when they are moved downwards. The so-called
`reference' buoyancy profile
$\beta^{*}:\mathbb{R}\rightarrow\mathbb{R}$ in available potential
energy theory is a volume-preserving function of height that
minimises potential energy \citep[see
e.g.][]{WinKjfm1995a}. `Volume-preserving' in this context means
that the mapping $\beta^{*}$ does not change the marginal
distribution of buoyancy, which, in the present context,
means that $\mathbb{E}[g\circ\beta^{*}(Z_{t})]=\mathbb{E}[g(B_{t})]$ for any
observable $g$. 

The potential energy \eqref{eq:bz} is minimised by placing
positively buoyant parcels at the top of the domain and negatively
buoyant parcels at the bottom of the domain. The
required mapping $\beta^{*}$ is monotonic and corresponds to
\begin{equation}
  \beta^{*}:=F_{B}^{-1}\circ F_{Z},
  \label{eq:bs}
\end{equation}
\noindent where $F_{B}$ and $F_{Z}$ are the marginal cumulative
distribution functions for $B_{t}$ and $Z_{t}$, respectively. In
the field of optimal transport, $\beta^{*}$ minimises the expected
squared distance between two variables and has been known for a
long time \citep{KnoMjot1984a, MccRdmj1995a}. The reference state
allows the available potential energy to be computed as the
difference between the actual potential energy and the (minimal)
potential energy associated with the reference state:
\begin{equation}
  \mathbb{E}\left[\beta^{*}(Z_{t})Z_{t}\right]-\mathbb{E}\left[B_{t}Z\right]=\int\limits_{\mathcal{Y}}(\beta^{*}(z)-b)zf_{\vc{Y}}(\vc{y},t)\rd \vc{y}\geq 0.
\end{equation}

A key related quantity in stratified turbulence is the
horizontally-averaged vertical buoyancy flux
$\mathbb{E}_{Z}[W_{t}B_{t}]$, which is responsible for the
reversible conversion of available potential energy into kinetic
energy \citep[see, for example,][]{CauCprf2020a}. In the averaged
Boussinesq equations, the horizontally-averaged vertical buoyancy
flux requires closure, but from the perspective of the joint
density for $B_{t}$ and $W_{t}$, conditional on $Z_{t}$, it is
known exactly in terms of independent and dependent variables:

\begin{equation}
  \mathbb{E}_{Z}[W_{t}B_{t}]=\iint\limits_{\mathbb{R}^{2}}wbf_{BW|Z}(b,w,z,t)\rd w\rd b,\quad\quad\forall z: f_{Z}(z)\neq 0,
\end{equation}

\noindent which suggests that the evolution equations for $f_{\vc{Y}}$ might be
a useful perspective to adopt for modelling stratified turbulence and
making use of available energetics in a prognostic capacity.

\section{Conclusions}

We have derived the equation that governs the evolution of the
joint probability distribution of a set of local flow observables
drawn indiscriminately from a control volume. Since the joint
distributions pertain to the entire contents of a control volume,
they are different from the local, pointwise distributions that
one studies using traditional probability density methods. In
particular, the boundary conditions applied to the control volume
appear as internal forcing terms in probability space. The framework
can account naturally for heterogeneous and non-stationary stochastic
forcing from the interior or boundary of a domain (see example
\S\ref{sec:1D_Diffusion}). Terms that require closure appear as
conditional expectations over the entire control volume. Although
no assumptions were made regarding homogeneity of the control
volume or boundary conditions, we expect closure of the equations
to be more challenging for heterogeneous fields.

The governing equations were derived from general transport
equations, making the approach applicable to a far wider range of
bulk modelling problems in the mathematical, physical and
engineering sciences than was originally anticipated, such as the
motion of fluids, contaminant transport and heat transport in
gases, liquids and solids. It is tempting to speculate that
aspects of the framework might find broader appeal still, in
problems that involve monitoring distributions within control
volumes such as species in ecological modelling.

The approach we describe is useful because it can be applied to
control volumes of any shape, size and dimension, and readily
accommodates an arbitrary number of flow observables. As
demonstrated in \S\ref{sec:bou}, coordinates can be regarded as
observables, which means that the equations derived in
\S\ref{sec:gov} can be applied without modification to slices of a
domain parameterised by a coordinate.  In this regard, we expect
the incorporation of geopotential height to be useful in
addressing problems that involve gravitational potential energy.
Indeed, this work illustrates that the equations governing global
constructions of available and background potential energy
\citep{WinKjfm1995a} are specific examples of the more general
principles that determine the evolution of probability
distributions.

Probability density methods highlight the distinct roles played by
stirring and mixing in transport problems
\citep{VilEafm2020a}. Stirring due to advection within a control
volume has no effect on a quantity's probability
distribution. Only when advection occurs across a control volume's
boundary or when a domain is decomposed into coordinate slices
(see example \S\ref{sec:bou}) does stirring enter the governing
equation for the probability distribution. The primacy of
irreversible mixing, on the other hand, emerges as the main
challenge of obtaining closure of the system of equations. Indeed,
as identified by Pope in the context of local PDF methods
\citep{PopSboo2000a}, modelling molecular diffusion becomes the
central issue in local probability density methods. For
applications, further work can address closures for the negative
diffusion term by reinterpreting existing closure schemes in an
integrated/volumetric sense.

This work has not considered moving boundaries, but the authors
see no obvious reason that the approach could not be extended to
such cases, provided that appropriate care is taken in handling
the boundary fluxes \cite{ReeMjfm2021a}.

{\small \section*{Acknowledgements}

This work was supported by the Engineering and
Physical Sciences Research Council [grant number EP/V033883/1]
as part of the [D$^{*}$]stratify project.}

\appendix

\section{The transformation of probability density}
\label{sec:pushforward}

A function $\varphi_{t}:\mathcal{X}\rightarrow\mathcal{Y}$ that
assigns to coordinates $\vc{X}\in\mathcal{X}$ the value of the
dependent variables $\vc{Y}_{t}\in\mathcal{Y}$, transforms a
distribution over $\mathcal{X}$ into a distribution over
$\mathcal{Y}$. In particular the joint probability density
$f_{\vc{Y}}:\mathcal{Y}\rightarrow \mathbb{R}$ (if it exists)
must provide a consistent means of computing the expectation of an
observable $g:\mathcal{Y}\rightarrow \mathbb{R}$:

\begin{equation}
  \mathbb{E}[g]=\int\limits_{\mathcal{Y}}g(\vc{y})f_{\vc{Y}}(\vc{y})\rd\vc{y}=\frac{1}{\mu^{d}(\mathcal{X})}\int\limits_{\mathcal{X}}g(\varphi_{t}(\vc{x}))\rd\vc{x}.
  \label{eq:pushforward}
\end{equation}

\noindent Notice that if $\varphi_{t}(\vc{x})$ is constant over
subsets of $\mathcal{X}$ of nonzero measure then such a density
$f_{\vc{Y}}$ does not exist. In that case $f_{\vc{Y}}\rd \vc{y}$
would need to be replaced with the pushforward measure
$\nu(\rd\vc{y})$ \citep{BogVboo2007a, KaiCboo2001a} to
account for individual values $\vc{y}$ that occur over subsets of
$\mathcal{X}$ of finite size (referred to as `atoms' of
$\nu$ \citep{KaiCboo2001a}). If, in
the cases that the density $f_{\vc{Y}}$ does exist, we pick $g=\mathds{1}_{C(\vc{y})}$ as the indicator function
for the sub-codomain
$C(\vc{y}):\{\vc{y}\in\mathcal{Y}:y_{i}'\leq y_{i}\}$
\noindent then

\begin{equation}
 f_{\vc{Y}}(\vc{y},t)=\frac{\partial^{n}}{\partial y_{1}\ldots\partial y_{n}}\int\limits_{\varphi_{t}^{-1}(C(\vc{y}))}\frac{\rd\vc{x}}{\mu^{d}(\vc{X})},
\end{equation}

\noindent where
$\varphi^{-1}_{t}(C(\vc{y})):=\{\vc{X}\in\mathcal{X}:\varphi_{t}(\vc{X})\in
C(\vc{y})\}$ is the preimage of $\varphi_{t}$, which accounts for the
possibility that several distinct values $\vc{X}$ might map to the
same value $\vc{Y}$. If, however, $\varphi_{t}$ is invertible when
restricted to a subdomain $D\subset\mathcal{X}$, then
\begin{equation}
f_{\vc{Y}|D}(\vc{y},t)=\frac{1}{\mu^{d}(D)|J_{t}(\vc{y})|},
\label{eq:fD}
\end{equation}
\noindent where $f_{\vc{Y}|D}$ is the probability density
conditional on $\vc{X}\in D$, $\mu^{d}(D)$ is the size/volume
of $D$ and $J_{t}(\vc{y})$ is the Jacobian matrix
$\partial\varphi_{t}/\partial{\vc{X}}$ (expressed in terms of
$\vc{y}$). The global density $f_{\vc{Y}}$ can be calculated from 
\eqref{eq:fD} using Bayes' theorem to account for all such
contributions in a partition $\mathscr{P}$ of $\mathcal{X}$:

\begin{equation}
f_{\vc{Y}}(\vc{y},t)=\sum_{D\in\mathscr{P}}f_{\vc{Y}|D}(\vc{y},t)\frac{\mu^{d}(D)}{\mu^{d}(\mathcal{X})},
\label{eq:fYsum}
\end{equation}

\noindent where the ratio
$\mu^{d}(D)/\mu^{d}(\mathcal{X})$ effectively
accounts for the size, and therefore probability,
associated with $D$ relative to $\mathcal{X}$.

\section{The Lorenz (1963) model}
\label{sec:lorenz}

Lorenz's (1963) model for convection \cite{LorEjas1963a} is a
truncated solution of the Boussinesq equations for which the
vertical velocity $Y^{1}_{t}$ and buoyancy field $Y^{2}_{t}$,
relative to linear conduction, on a horizontally periodic domain
$\mathcal{X}:=[0,2\pi/k)\times [0,1]\ni \vc{X}:=(X_{1},X_{2})$ are
\begin{align}
  Y_{1}&=\varphi_{t}^{1}(X_{1},X_{2}):=\frac{\sqrt{2}}{\pi}\left(k^{2}+\pi^{2}\right)a_{1}(t)\mathrm{cos}(kX_{1})\mathrm{sin}(\pi X_{2}),\label{eq:phi_a}
\\
  Y_{2}&=\varphi_{t}^{2}(X_{1},X_{2}):=\frac{\sqrt{2}}{\pi r}a_{2}(t)\mathrm{cos}(kX_{1})\mathrm{sin}(\pi X_{2})-\frac{1}{\pi r}a_{3}(t)\mathrm{sin}(2\pi X_{2}),
\label{eq:phi_b}
\end{align}
\noindent where $r$ is a renormalised Rayleigh number. The amplitudes $\vc{a}:=(a_{1},a_{2},a_{3})^{\top}$ evolve in time according to
\begin{equation}
  \od{a_{1}}{t}=s(a_{2}-a_{1}),\quad
  \od{a_{2}}{t}=ra_{1}-a_{2}-a_{1}a_{3},\quad
  \od{a_{3}}{t}=a_{1}a_{2}-ba_{3},
  \label{eq:lorenz}
\end{equation}

\noindent where $b:=4\pi^{2}(k^{2}+\pi^{2})^{-1}$ characterises
the aspect ratio of the domain and $s$ is the Prandtl number. To
construct the density
$f_{\vc{Y}}(-,t):\mathcal{Y}\rightarrow \mathbb{R}$ it is
sufficient to consider the subdomains
$D_{1}:=(0,\pi/k]\times [0,1/4)$ and
$D_{2}:=(0,\pi/k]\times (1/4,1/2]$ from which the fields in other
parts of the global domain can be readily constructed using
symmetry arguments (cf. the arguments leading to \eqref{eq:fYsum}
in appendix \ref{sec:pushforward}, which involve partitioning a
domain into parts for which $\varphi_{t}$ is invertible).

Over $D_{1}$ and $D_{2}$ the Jacobian
$J:=\partial \varphi_{t}/\partial \vc{X}$ is non-zero,
which means that $\varphi_{t}$ has a single-valued inverse when its
domain is restricted to $D_{1}$ and $D_{2}$ that can be found by
manipulation of \eqref{eq:phi_b}.

\bibliographystyle{rs}
{\small
\bibliography{main}
}

\end{document}

%% file: example1d.tex
\definecolor{c3c008c}{RGB}{60,0,140}
\definecolor{caa1515}{RGB}{170,21,21}
\definecolor{c050302}{RGB}{5,3,2}

\def \globalscale {1.000000}
\begin{tikzpicture}[y=1cm, x=1cm, yscale=\globalscale,xscale=\globalscale, every node/.append style={scale=\globalscale}, inner sep=0pt, outer sep=0pt]
  \path[draw=black,line cap=butt,line join=miter,line width=0.0163cm] (6.1147, 25.6784) -- (12.9776, 25.6784);

  \path[draw=black,line cap=butt,line join=miter,line width=0.0353cm] (6.1147, 23.2062) -- (12.2685, 23.2062);

  \path[draw=black,line cap=butt,line join=miter,line width=0.0163cm] (13.6072, 27.8784) -- (13.6072, 23.4783);

  \path[draw=black,line cap=butt,line join=miter,line width=0.0163cm] (13.6072, 25.6784) -- (16.2961, 25.6784);

  \path[draw=black,line cap=butt,line join=miter,line width=0.0163cm] (6.0515, 27.7) -- (6.121, 27.8784) -- (6.1938, 27.7062);

  \path[draw=black,line cap=butt,line join=miter,line width=0.0163cm] (12.8869, 25.7367).. controls (12.8963, 25.7336) and (12.9909, 25.6754) .. (12.9909, 25.6754) -- (12.8791, 25.6196);

  \path[draw=black,line cap=butt,line join=miter,line width=0.0163cm] (16.1676, 25.7341) -- (16.2961, 25.6784) -- (16.1684, 25.6204);

  \path[draw=black,line cap=butt,line join=miter,line width=0.0163cm] (13.5441, 27.7221) -- (13.6072, 27.8784) -- (13.6732, 27.7243);

  \path[draw=c3c008c,line cap=butt,line join=miter,line width=0.0326cm] (10.933, 24.3658).. controls (11.7235, 24.4041) and (11.0161, 27.1174) .. (11.8021, 27.1256).. controls (11.8021, 27.1256) and (12.1467, 27.1644) .. (12.2414, 26.5174);

  \path[draw=c3c008c,line cap=butt,line join=miter,line width=0.0326cm] (6.1187, 23.8442).. controls (6.1187, 23.8442) and (6.5284, 23.7336) .. (6.8683, 24.6157).. controls (7.2082, 25.4979) and (7.2931, 26.7992) .. (7.674, 26.7869).. controls (8.0548, 26.7747) and (8.1231, 25.1592) .. (8.4387, 25.1578).. controls (8.7543, 25.1565) and (8.5874, 26.1147) .. (9.0639, 26.1124).. controls (9.6238, 26.1096) and (9.3764, 24.3849) .. (10.0319, 24.3672);

  \path[draw=black,line cap=butt,line join=miter,line width=0.0098cm] (11.2801, 27.1242) -- (12.2269, 27.1242);

  \path[draw=black,line cap=butt,line join=miter,line width=0.0098cm] (7.3569, 26.7842) -- (7.9805, 26.7842);

  \path[draw=black,line cap=butt,line join=miter,line width=0.0098cm] (8.6884, 26.1042) -- (9.4076, 26.1042);

  \path[draw=black,line cap=butt,line join=miter,line width=0.0098cm] (8.1436, 25.1491) -- (8.7566, 25.1491);

  \path[draw=black,line cap=butt,line join=miter,line width=0.0098cm] (9.4777, 24.3608) -- (10.1876, 24.3608);

  \path[draw=black,line cap=butt,line join=miter,line width=0.0098cm] (10.8118, 24.3623) -- (11.5216, 24.3623);

  \path[draw=black,line cap=butt,line join=miter,line width=0.0098cm] (6.1976, 23.8524) -- (6.5936, 23.8524);

  \path[draw=black,line cap=butt,line join=miter,line width=0.0163cm] (6.121, 27.8784) -- (6.1278, 23.4783);

  \path[draw=caa1515,line cap=butt,line join=miter,line width=0.0326cm] (14.8686, 27.118).. controls (14.2138, 27.1347) and (13.7084, 27.0782) .. (13.6942, 26.9689).. controls (13.6091, 26.7203) and (14.8892, 26.7857) .. (15.0916, 26.7839);

  \path[draw=caa1515,line cap=butt,line join=miter,line width=0.0326cm] (15.0461, 26.7705).. controls (14.5033, 26.7723) and (13.8396, 26.7304) .. (13.7162, 26.5989).. controls (13.6308, 26.5079) and (13.6564, 26.2716) .. (13.7769, 26.2265).. controls (14.1087, 26.1024) and (14.7675, 26.107) .. (14.7675, 26.107);

  \path[draw=caa1515,line cap=butt,line join=miter,line width=0.0326cm] (14.7666, 26.0922).. controls (14.7666, 26.0922) and (13.9423, 26.0807) .. (13.7698, 25.8796).. controls (13.6215, 25.7422) and (13.6256, 25.3887) .. (13.8225, 25.2976).. controls (14.304, 25.0749) and (14.7415, 25.1827) .. (15.3071, 25.1541);

  \path[draw=caa1515,line cap=butt,line join=miter,line width=0.0326cm] (15.2717, 25.1463).. controls (15.2717, 25.1463) and (13.8507, 25.1099) .. (13.7092, 24.9405).. controls (13.6655, 24.8882) and (13.6378, 24.6327) .. (13.7799, 24.4951).. controls (13.922, 24.3576) and (14.7969, 24.3825) .. (15.8925, 24.3659);

  \path[draw=caa1515,line cap=butt,line join=miter,line width=0.0326cm] (15.8974, 24.36).. controls (15.2285, 24.3414) and (14.2499, 24.4098) .. (13.754, 24.2403).. controls (13.6574, 24.2074) and (13.624, 24.0598) .. (13.7494, 23.9699).. controls (13.8748, 23.88) and (14.8233, 23.8456) .. (14.8233, 23.8456);

  \path[draw=c3c008c,fill=c3c008c,line cap=butt,line join=miter,line width=0.0326cm,dash pattern=on 0.1304cm off 0.1304cm] (10.0433, 24.361) -- (10.9326, 24.361);

  \node[line width=0.0244cm,anchor=south west] (text2) at (5.568, 27.6072){$Y$};

  \node[line width=0.0244cm,anchor=south west] (text2-7) at (12.7491, 25.2216){$X$};

  \node[line width=0.0244cm,anchor=south west] (text2-7-6) at (12.2563, 25.1344){$1$};

  \node[line width=0.0244cm,anchor=south west] (text2-7-6-1) at (5.7489, 25.3378){$0$};

  \node[line width=0.0244cm,anchor=south west] (text2-7-6-8) at (10.3888, 25.1343){$\mu$};

  \node[line width=0.0244cm,anchor=south west] (text2-7-2) at (16.348, 23.3704){$\dfrac{\delta(y-y_{0})}{\mu}$};

  \node[line width=0.0244cm,anchor=south west] (text2-7-2-0) at (8.8078, 24.2224){$y_{0}$};

  \node[line width=0.0244cm,anchor=south west] (text2-7-2-0-9) at (13.1382, 24.2285){$y_{0}$};

  \node[line width=0.0244cm,anchor=south west] (text2-7-7) at (6.2977, 25.8453){$X(\omega)$};

  \node[line width=0.0244cm,anchor=south west] (text2-7-7-9) at (5.1124, 24.1523){$Y(\omega)$};

  \node[line width=0.0244cm,anchor=south west] (text2-7-7-9-7) at (6.6224, 22.7782){$\omega$};

  \node[line width=0.0244cm,anchor=south west] (text2-7-9) at (12.3542, 23.028){$\Omega$};

  \node[line width=0.0244cm,anchor=south west] (text2-7-0) at (13.0731, 27.7537){$y$};

  \node[line width=0.0244cm,anchor=south west] (text2-7-0-8) at (15.7933, 25.2049){$f_{Y}(y)$};

  \path[draw=black,line cap=butt,line join=miter,line width=0.0176cm,dash pattern=on 0.0706cm off 0.0706cm] (6.7665, 23.2195) -- (6.7665, 24.2092) -- (6.7665, 25.6752);

  \path[draw=black,line cap=butt,line join=miter,line width=0.0176cm,dash pattern=on 0.0706cm off 0.0706cm] (6.7602, 24.352) -- (6.1424, 24.352);

  \path[draw=black,line cap=butt,line join=miter,line width=0.0176cm] (6.335, 24.254) -- (6.1953, 24.3492) -- (6.3445, 24.4539);

  \path[draw=black,line cap=butt,line join=miter,line width=0.0176cm] (6.6809, 25.542) -- (6.7665, 25.6782) -- (6.8543, 25.5563);

  \path[draw=black,fill=c050302,line cap=round,line join=round,line width=0.0107cm] (13.6065, 24.367) circle (0.0449cm);

  \path[draw=black,fill=c050302,line cap=round,line join=round,line width=0.0176cm,shift={(0.1454, -0.0)}] (12.2597, 25.8286) -- (12.2597, 25.5007);

  \path[draw=black,fill=c050302,line cap=round,line join=round,line width=0.0176cm,shift={(-2.2738, -0.223)}] (12.2597, 25.8286) -- (12.2597, 25.5007);

  \path[draw=black,fill=c050302,line cap=round,line join=round,line width=0.0176cm,shift={(-1.2102, -0.2195)}] (12.2597, 25.8286) -- (12.2597, 25.5007);

  \path[draw=black,fill=c050302,line cap=round,line join=round,line width=0.0176cm,shift={(0.0, 0.1677)}] (9.9789, 25.2858) -- (11.0465, 25.2858);

  \path[draw=caa1515,fill=caa1515,line cap=round,line join=round,line width=0.0282cm,dash pattern=on 0.2258cm off 0.1129cm] (15.9559, 24.3607) -- (17.781, 24.3607);

  \node[line width=0.0244cm,anchor=south west] (text2-7-6-8-5) at (8.9836, 26.6372){$\varphi(X)$};

  \path[draw=black,line cap=butt,line join=miter,line width=0.0176cm,dash pattern=on 0.0706cm off 0.0706cm] (8.0327, 26.47).. controls (8.0327, 26.47) and (8.3246, 26.7453) .. (8.8668, 26.787);

\end{tikzpicture}

%% file: diag.tex
\definecolor{c5b5b5b}{RGB}{91,91,91}
\definecolor{cf0e0dd}{RGB}{240,224,221}
\definecolor{cd9aea4}{RGB}{217,174,164}
\definecolor{ce4c6c0}{RGB}{228,198,192}
\definecolor{ce6cdc8}{RGB}{230,205,200}
\definecolor{ce3c4bc}{RGB}{227,196,188}
\definecolor{cd29f94}{RGB}{210,159,148}
\definecolor{ce7e7eb}{RGB}{231,231,235}
\definecolor{c9e9eae}{RGB}{158,158,174}
\definecolor{c242060}{RGB}{36,32,96}
\definecolor{c282463}{RGB}{40,36,99}

\def \globalscale {1.000000}
\begin{tikzpicture}[y=1cm, x=1cm, yscale=\globalscale,xscale=\globalscale, every node/.append style={scale=\globalscale}, inner sep=0pt, outer sep=0pt]
  \path[draw=c5b5b5b,fill=cf0e0dd,even odd rule,line cap=butt,line join=miter,line width=0.0176cm] (5.4456, 23.9872).. controls (5.2584, 24.3023) and (5.2083, 24.6958) .. (5.3106, 25.0477).. controls (5.413, 25.3997) and (5.6663, 25.7049) .. (5.9933, 25.8705).. controls (6.3143, 26.0329) and (6.6876, 26.06) .. (7.0466, 26.037).. controls (7.4057, 26.014) and (7.7607, 25.9443) .. (8.1199, 25.9252).. controls (8.5444, 25.9025) and (8.9719, 25.9507) .. (9.3942, 25.9025).. controls (9.6054, 25.8784) and (9.8153, 25.8297) .. (10.0077, 25.7393).. controls (10.2001, 25.649) and (10.3747, 25.5153) .. (10.4962, 25.341).. controls (10.6186, 25.1654) and (10.6839, 24.9525) .. (10.6921, 24.7386).. controls (10.7003, 24.5247) and (10.6526, 24.3102) .. (10.5649, 24.115).. controls (10.3893, 23.7246) and (10.0609, 23.4198) .. (9.6973, 23.1939).. controls (9.0607, 22.7984) and (8.2885, 22.6187) .. (7.5452, 22.7147).. controls (6.8019, 22.8108) and (6.0943, 23.1863) .. (5.6149, 23.7624).. controls (5.5548, 23.8345) and (5.4983, 23.9096) .. (5.4456, 23.9872) -- cycle;

  \path[draw=c5b5b5b,fill=cf0e0dd,even odd rule,line cap=butt,line join=miter,line width=0.0176cm] (5.3977, 23.8244).. controls (5.2104, 24.1394) and (5.1603, 24.5329) .. (5.2627, 24.8849).. controls (5.365, 25.2368) and (5.6183, 25.5421) .. (5.9453, 25.7076).. controls (6.2663, 25.8701) and (6.6396, 25.8972) .. (6.9987, 25.8742).. controls (7.3577, 25.8512) and (7.7127, 25.7814) .. (8.072, 25.7623).. controls (8.4965, 25.7397) and (8.9239, 25.7879) .. (9.3463, 25.7397).. controls (9.5575, 25.7156) and (9.7674, 25.6669) .. (9.9597, 25.5765).. controls (10.1521, 25.4861) and (10.3267, 25.3524) .. (10.4483, 25.1781).. controls (10.5707, 25.0025) and (10.636, 24.7896) .. (10.6442, 24.5757).. controls (10.6524, 24.3618) and (10.6047, 24.1474) .. (10.5169, 23.9521).. controls (10.3414, 23.5617) and (10.0129, 23.257) .. (9.6493, 23.031).. controls (9.0128, 22.6355) and (8.2406, 22.4558) .. (7.4973, 22.5519).. controls (6.754, 22.6479) and (6.0464, 23.0234) .. (5.5669, 23.5995).. controls (5.5069, 23.6716) and (5.4504, 23.7467) .. (5.3977, 23.8244) -- cycle;

  \path[draw=c5b5b5b,fill=cf0e0dd,even odd rule,line cap=butt,line join=miter,line width=0.0176cm] (5.3583, 23.6481).. controls (5.171, 23.9632) and (5.121, 24.3567) .. (5.2233, 24.7086).. controls (5.3256, 25.0605) and (5.5789, 25.3658) .. (5.9059, 25.5313).. controls (6.2269, 25.6938) and (6.6002, 25.7209) .. (6.9593, 25.6979).. controls (7.3183, 25.6749) and (7.6733, 25.6052) .. (8.0326, 25.586).. controls (8.4571, 25.5634) and (8.8846, 25.6116) .. (9.3069, 25.5634).. controls (9.5181, 25.5393) and (9.728, 25.4906) .. (9.9204, 25.4002).. controls (10.1127, 25.3098) and (10.2873, 25.1762) .. (10.4089, 25.0019).. controls (10.5313, 24.8263) and (10.5966, 24.6134) .. (10.6048, 24.3995).. controls (10.613, 24.1856) and (10.5653, 23.9711) .. (10.4775, 23.7759).. controls (10.302, 23.3855) and (9.9736, 23.0807) .. (9.61, 22.8548).. controls (8.9734, 22.4592) and (8.2012, 22.2796) .. (7.4579, 22.3756).. controls (6.7146, 22.4717) and (6.007, 22.8472) .. (5.5275, 23.4233).. controls (5.4675, 23.4954) and (5.411, 23.5705) .. (5.3583, 23.6481) -- cycle;

  \path[draw=c5b5b5b,fill=cf0e0dd,even odd rule,line cap=butt,line join=miter,line width=0.0176cm] (5.3134, 23.4662).. controls (5.1261, 23.7813) and (5.0761, 24.1748) .. (5.1784, 24.5267).. controls (5.2807, 24.8786) and (5.534, 25.1839) .. (5.861, 25.3494).. controls (6.182, 25.5119) and (6.5553, 25.539) .. (6.9144, 25.516).. controls (7.2734, 25.493) and (7.6284, 25.4233) .. (7.9877, 25.4041).. controls (8.4122, 25.3815) and (8.8397, 25.4297) .. (9.262, 25.3815).. controls (9.4732, 25.3574) and (9.6831, 25.3087) .. (9.8755, 25.2183).. controls (10.0678, 25.1279) and (10.2424, 24.9943) .. (10.364, 24.8199).. controls (10.4864, 24.6444) and (10.5517, 24.4314) .. (10.5599, 24.2176).. controls (10.5681, 24.0037) and (10.5204, 23.7892) .. (10.4326, 23.594).. controls (10.2571, 23.2035) and (9.9287, 22.8988) .. (9.5651, 22.6729).. controls (8.9285, 22.2773) and (8.1563, 22.0976) .. (7.413, 22.1937).. controls (6.6697, 22.2897) and (5.9621, 22.6653) .. (5.4826, 23.2413).. controls (5.4226, 23.3135) and (5.3661, 23.3885) .. (5.3134, 23.4662) -- cycle;

  \path[draw=black,even odd rule,line cap=butt,line join=miter,line width=0.0353cm] (5.2613, 23.2896).. controls (5.074, 23.6047) and (5.0239, 23.9982) .. (5.1263, 24.3501).. controls (5.2286, 24.7021) and (5.4819, 25.0073) .. (5.8089, 25.1728).. controls (6.1299, 25.3353) and (6.5032, 25.3624) .. (6.8623, 25.3394).. controls (7.2213, 25.3164) and (7.5763, 25.2467) .. (7.9356, 25.2275).. controls (8.3601, 25.2049) and (8.7875, 25.2531) .. (9.2099, 25.2049).. controls (9.4211, 25.1808) and (9.631, 25.1321) .. (9.8233, 25.0417).. controls (10.0157, 24.9513) and (10.1903, 24.8177) .. (10.3119, 24.6434).. controls (10.4343, 24.4678) and (10.4996, 24.2549) .. (10.5078, 24.041).. controls (10.516, 23.8271) and (10.4683, 23.6126) .. (10.3805, 23.4174).. controls (10.205, 23.027) and (9.8766, 22.7222) .. (9.513, 22.4963).. controls (8.8764, 22.1007) and (8.1042, 21.9211) .. (7.3609, 22.0171).. controls (6.6176, 22.1132) and (5.91, 22.4887) .. (5.4305, 23.0648).. controls (5.3705, 23.1369) and (5.314, 23.212) .. (5.2613, 23.2896) -- cycle;

  \path[draw=black,even odd rule,line cap=butt,line join=miter,line width=0.0176cm] (5.0749, 23.9259).. controls (5.6512, 24.1281) and (6.1472, 24.1287) .. (6.7516, 24.2162).. controls (7.2205, 24.284) and (7.6949, 24.3166) .. (8.1685, 24.3031).. controls (8.4406, 24.2953) and (8.7125, 24.2724) .. (8.9846, 24.2799).. controls (9.2567, 24.2873) and (9.5318, 24.3265) .. (9.7805, 24.4372).. controls (9.9473, 24.5116) and (10.1001, 24.6176) .. (10.2281, 24.7479);

  \path[draw=black,even odd rule,line cap=butt,line join=miter,line width=0.0176cm] (6.1066, 22.4781).. controls (6.3522, 22.8262) and (6.8305, 23.2642) .. (7.2034, 23.4701).. controls (7.3944, 23.5756) and (7.5991, 23.6593) .. (7.8137, 23.6984).. controls (8.0283, 23.7375) and (8.2535, 23.7307) .. (8.4598, 23.66).. controls (8.6751, 23.5862) and (8.8607, 23.4474) .. (9.0502, 23.3215).. controls (9.2398, 23.1955) and (9.26, 23.14) .. (9.4865, 23.118).. controls (9.6705, 23.1002) and (9.7558, 23.202) .. (9.9079, 23.307).. controls (10.0601, 23.4119) and (10.121, 23.7246) .. (10.4402, 23.572);

  \path[draw=black,even odd rule,line cap=butt,line join=miter,line width=0.0176cm] (6.0757, 25.2722).. controls (5.9479, 24.9714) and (5.9246, 24.875) .. (5.6662, 24.6676).. controls (5.5614, 24.5835) and (5.3039, 24.4713) .. (5.1696, 24.4749);

  \path[draw=black,even odd rule,line cap=butt,line join=miter,line width=0.0176cm] (7.6993, 22.5009).. controls (7.6459, 22.5296) and (7.6029, 22.5773) .. (7.5799, 22.6334).. controls (7.5569, 22.6895) and (7.554, 22.7536) .. (7.5718, 22.8115).. controls (7.5858, 22.8571) and (7.6123, 22.8985) .. (7.6469, 22.9313).. controls (7.6815, 22.964) and (7.7242, 22.988) .. (7.7698, 23.0017).. controls (7.861, 23.0292) and (7.9624, 23.0146) .. (8.0466, 22.97).. controls (8.0984, 22.9425) and (8.1449, 22.9036) .. (8.1763, 22.8542).. controls (8.2078, 22.8048) and (8.2236, 22.7444) .. (8.2149, 22.6865).. controls (8.2092, 22.6485) and (8.1932, 22.6123) .. (8.1702, 22.5816).. controls (8.1472, 22.5509) and (8.1173, 22.5256) .. (8.0838, 22.5067).. controls (8.017, 22.4689) and (7.9376, 22.4574) .. (7.8611, 22.4631).. controls (7.8056, 22.4672) and (7.7508, 22.48) .. (7.6993, 22.5009) -- cycle;

  \path[draw=black,even odd rule,line cap=butt,line join=miter,line width=0.0176cm] (8.0307, 22.0033).. controls (8.2471, 22.1098) and (8.309, 22.3037) .. (8.7798, 22.4501).. controls (8.9885, 22.515) and (9.2081, 22.5484) .. (9.4061, 22.641).. controls (9.5128, 22.691) and (9.612, 22.7553) .. (9.7145, 22.8135).. controls (9.8169, 22.8717) and (9.9243, 22.9244) .. (10.0398, 22.9479);

  \path[fill=cd9aea4,fill opacity=0.6329,line cap=round,line join=round,line width=0.0058cm] (7.7693, 22.4918).. controls (7.6066, 22.5384) and (7.5329, 22.6931) .. (7.6024, 22.8422).. controls (7.6857, 23.0211) and (7.9467, 23.054) .. (8.1103, 22.9061).. controls (8.1739, 22.8488) and (8.1982, 22.7965) .. (8.1982, 22.7177).. controls (8.1982, 22.5798) and (8.0849, 22.4859) .. (7.9114, 22.4798).. controls (7.8582, 22.4779) and (7.8005, 22.4828) .. (7.7693, 22.4918) -- cycle;

  \path[fill=cf0e0dd,line cap=round,line join=round,line width=0.0058cm] (7.4019, 22.0348).. controls (7.1607, 22.0648) and (6.911, 22.1256) .. (6.6854, 22.2093).. controls (6.5074, 22.2754) and (6.1364, 22.4636) .. (6.1364, 22.4878).. controls (6.1364, 22.5249) and (6.4411, 22.8604) .. (6.6202, 23.0203).. controls (7.0093, 23.368) and (7.3508, 23.5693) .. (7.7028, 23.6584).. controls (8.0399, 23.7438) and (8.3244, 23.7186) .. (8.6073, 23.5784).. controls (8.7372, 23.514) and (8.7881, 23.4821) .. (9.1149, 23.2601).. controls (9.3045, 23.1313) and (9.3708, 23.1062) .. (9.5211, 23.1064).. controls (9.6261, 23.1065) and (9.6889, 23.1318) .. (9.8209, 23.2269).. controls (9.9358, 23.3097) and (9.9762, 23.3482) .. (10.0633, 23.4578).. controls (10.1261, 23.5369) and (10.1894, 23.5847) .. (10.245, 23.5951).. controls (10.2905, 23.6037) and (10.3706, 23.5925) .. (10.3957, 23.5742).. controls (10.4131, 23.5615) and (10.4129, 23.5571) .. (10.392, 23.5042).. controls (10.3065, 23.2873) and (10.0837, 22.9563) .. (10.023, 22.9561).. controls (9.9666, 22.9559) and (9.7892, 22.8749) .. (9.6055, 22.7655).. controls (9.3961, 22.6408) and (9.3048, 22.6048) .. (8.9829, 22.5196).. controls (8.7396, 22.4552) and (8.6497, 22.4241) .. (8.52, 22.3594).. controls (8.4093, 22.3042) and (8.3551, 22.2672) .. (8.2206, 22.1552).. controls (8.0523, 22.0151) and (8.0798, 22.0236) .. (7.7837, 22.0212).. controls (7.615, 22.0199) and (7.4844, 22.0245) .. (7.4019, 22.0348) -- cycle(8.0397, 22.4731).. controls (8.1092, 22.4951) and (8.1987, 22.5822) .. (8.2177, 22.6463).. controls (8.2459, 22.7419) and (8.222, 22.8313) .. (8.1478, 22.9077).. controls (8.0379, 23.0208) and (7.8601, 23.0592) .. (7.7291, 22.9982).. controls (7.5357, 22.9082) and (7.4867, 22.6846) .. (7.6292, 22.5421).. controls (7.6713, 22.5) and (7.6937, 22.4873) .. (7.7513, 22.4725).. controls (7.8682, 22.4424) and (7.943, 22.4426) .. (8.0397, 22.4731) -- cycle;

  \path[fill=ce4c6c0,line cap=round,line join=round,line width=0.0041cm] (8.1303, 22.0493).. controls (8.143, 22.0592) and (8.1986, 22.1043) .. (8.2537, 22.1495).. controls (8.48, 22.335) and (8.6169, 22.3984) .. (9.0174, 22.503).. controls (9.3148, 22.5806) and (9.394, 22.612) .. (9.6047, 22.7357).. controls (9.672, 22.7752) and (9.7624, 22.8256) .. (9.8056, 22.8477).. controls (9.8818, 22.8866) and (10.0058, 22.9333) .. (10.0132, 22.9259).. controls (10.019, 22.9202) and (9.874, 22.7858) .. (9.7856, 22.715).. controls (9.5058, 22.4906) and (9.1208, 22.2863) .. (8.7628, 22.172).. controls (8.589, 22.1166) and (8.3258, 22.0585) .. (8.1496, 22.0367) -- (8.1071, 22.0314) -- cycle;

  \path[fill=ce6cdc8,line cap=round,line join=round,line width=0.0041cm] (6.0087, 22.5639).. controls (5.7405, 22.7528) and (5.507, 22.9828) .. (5.3213, 23.2407).. controls (5.1997, 23.4096) and (5.1239, 23.6254) .. (5.0991, 23.8735) -- (5.0944, 23.9203) -- (5.1969, 23.9522).. controls (5.4954, 24.0449) and (5.6874, 24.0789) .. (6.3214, 24.1509).. controls (6.455, 24.1661) and (6.6511, 24.1903) .. (6.7572, 24.2046).. controls (7.1088, 24.252) and (7.3198, 24.2719) .. (7.625, 24.2864).. controls (7.831, 24.2962) and (8.2417, 24.292) .. (8.5044, 24.2774).. controls (8.7217, 24.2653) and (9.1504, 24.2718) .. (9.2719, 24.2889).. controls (9.4094, 24.3083) and (9.5382, 24.3371) .. (9.6422, 24.3716).. controls (9.8434, 24.4384) and (10.0494, 24.5597) .. (10.187, 24.6923) -- (10.2226, 24.7266) -- (10.2739, 24.6585).. controls (10.3337, 24.5792) and (10.4, 24.4501) .. (10.4318, 24.3512).. controls (10.4718, 24.2268) and (10.4831, 24.1439) .. (10.4823, 23.9809).. controls (10.4816, 23.8316) and (10.4715, 23.7545) .. (10.4371, 23.6363) -- (10.4244, 23.5925) -- (10.3811, 23.6073).. controls (10.3186, 23.6289) and (10.2397, 23.6273) .. (10.1888, 23.6036).. controls (10.1347, 23.5784) and (10.0956, 23.5414) .. (10.0121, 23.4365).. controls (9.9534, 23.3627) and (9.9263, 23.3363) .. (9.8566, 23.2848).. controls (9.6461, 23.1294) and (9.5892, 23.1105) .. (9.4199, 23.1401).. controls (9.3224, 23.1571) and (9.2784, 23.1765) .. (9.1776, 23.2471).. controls (8.7088, 23.575) and (8.5588, 23.6563) .. (8.3261, 23.7084).. controls (8.2544, 23.7245) and (8.2269, 23.7263) .. (8.0647, 23.7261).. controls (7.8971, 23.7259) and (7.876, 23.7243) .. (7.785, 23.7048).. controls (7.4215, 23.6272) and (7.1027, 23.4587) .. (6.7194, 23.1415).. controls (6.5883, 23.033) and (6.2865, 22.7286) .. (6.1904, 22.6078).. controls (6.1465, 22.5527) and (6.1068, 22.5067) .. (6.1022, 22.5057).. controls (6.0976, 22.5047) and (6.0555, 22.5309) .. (6.0087, 22.5639) -- cycle;

  \path[fill=ce3c4bc,line cap=round,line join=round,line width=0.0041cm] (5.0949, 23.9975).. controls (5.0951, 24.1137) and (5.1189, 24.2555) .. (5.1594, 24.3817) -- (5.1841, 24.4585) -- (5.2494, 24.4716).. controls (5.3795, 24.4977) and (5.5649, 24.5783) .. (5.6654, 24.6524).. controls (5.7401, 24.7074) and (5.8603, 24.8275) .. (5.9034, 24.8899).. controls (5.9493, 24.9565) and (5.9863, 25.0288) .. (6.0402, 25.1573) -- (6.0808, 25.2537) -- (6.1292, 25.267).. controls (6.215, 25.2905) and (6.3558, 25.3109) .. (6.4986, 25.3205).. controls (6.7145, 25.3351) and (6.9206, 25.3211) .. (7.4283, 25.2575).. controls (7.8193, 25.2085) and (7.7928, 25.2099) .. (8.4736, 25.2035).. controls (9.1326, 25.1974) and (9.181, 25.195) .. (9.3684, 25.1606).. controls (9.6086, 25.1165) and (9.8196, 25.0356) .. (9.9901, 24.922).. controls (10.0691, 24.8694) and (10.2015, 24.7557) .. (10.2015, 24.7405).. controls (10.2015, 24.7369) and (10.1676, 24.7057) .. (10.1263, 24.6711).. controls (9.8846, 24.4687) and (9.6231, 24.3604) .. (9.2593, 24.3119).. controls (9.1154, 24.2927) and (8.7495, 24.2893) .. (8.4273, 24.3042).. controls (7.8081, 24.3328) and (7.3768, 24.3132) .. (6.7495, 24.228).. controls (6.6434, 24.2136) and (6.4612, 24.1911) .. (6.3445, 24.1779).. controls (5.8032, 24.117) and (5.6115, 24.0874) .. (5.3726, 24.0274).. controls (5.3028, 24.0099) and (5.1406, 23.9621) .. (5.1061, 23.9489).. controls (5.0969, 23.9454) and (5.0949, 23.9543) .. (5.0949, 23.9975) -- cycle;

  \path[fill=cd29f94,line cap=round,line join=round,line width=0.0041cm] (5.2029, 24.4954).. controls (5.2029, 24.5088) and (5.2837, 24.6585) .. (5.319, 24.7106).. controls (5.441, 24.8902) and (5.616, 25.0468) .. (5.8045, 25.1449).. controls (5.8854, 25.187) and (5.986, 25.229) .. (6.0282, 25.2383) -- (6.0504, 25.2432) -- (6.0135, 25.1559).. controls (5.9135, 24.9198) and (5.8453, 24.827) .. (5.6685, 24.6865).. controls (5.5844, 24.6196) and (5.4176, 24.5391) .. (5.2954, 24.5066).. controls (5.2278, 24.4886) and (5.2029, 24.4855) .. (5.2029, 24.4954) -- cycle;

  \path[fill=ce7e7eb,even odd rule,line cap=,line width=0.0198cm] (13.686, 23.5084) ellipse (2.5119cm and 1.129cm);

  \path[fill=c9e9eae,fill opacity=0.7155,even odd rule,line cap=butt,line join=miter,line width=0.0265cm] (11.3161, 23.1292).. controls (11.2402, 23.2186) and (11.1953, 23.3338) .. (11.1906, 23.4509).. controls (11.185, 23.5908) and (11.2358, 23.7289) .. (11.3134, 23.8454).. controls (11.382, 23.9483) and (11.471, 24.0363) .. (11.5681, 24.1128).. controls (11.7144, 24.2281) and (11.8812, 24.3189) .. (12.0605, 24.3696).. controls (12.1531, 24.3958) and (12.2484, 24.4113) .. (12.3419, 24.434).. controls (12.4355, 24.4567) and (12.5285, 24.4873) .. (12.6087, 24.5405).. controls (12.6965, 24.5989) and (12.7648, 24.6816) .. (12.8328, 24.7622).. controls (12.9664, 24.9205) and (13.1083, 25.0785) .. (13.2908, 25.1766).. controls (13.3683, 25.2182) and (13.4534, 25.2485) .. (13.5413, 25.2524).. controls (13.6205, 25.2559) and (13.6997, 25.2379) .. (13.7729, 25.2077).. controls (13.8462, 25.1775) and (13.914, 25.1354) .. (13.9787, 25.0895).. controls (14.1048, 25.0) and (14.2215, 24.894) .. (14.3055, 24.7642).. controls (14.3417, 24.7082) and (14.3734, 24.6463) .. (14.4279, 24.6078).. controls (14.4821, 24.5696) and (14.5508, 24.5603) .. (14.6162, 24.5494).. controls (14.7564, 24.5258) and (14.8941, 24.4893) .. (15.0309, 24.4508).. controls (15.1298, 24.4229) and (15.2286, 24.3939) .. (15.3291, 24.3724).. controls (15.392, 24.359) and (15.4559, 24.3484) .. (15.5159, 24.3251).. controls (15.5758, 24.3018) and (15.6326, 24.2643) .. (15.6654, 24.2089).. controls (15.6728, 24.1964) and (15.6789, 24.183) .. (15.6825, 24.1689).. controls (15.6861, 24.1548) and (15.687, 24.1399) .. (15.6842, 24.1256).. controls (15.6796, 24.1024) and (15.6651, 24.0817) .. (15.6461, 24.0675).. controls (15.6272, 24.0533) and (15.6042, 24.0452) .. (15.5808, 24.0417).. controls (15.534, 24.0347) and (15.4866, 24.0457) .. (15.4398, 24.0527).. controls (15.3496, 24.066) and (15.2575, 24.0643) .. (15.1682, 24.0828).. controls (15.0068, 24.1162) and (14.8481, 24.1711) .. (14.7237, 24.2792).. controls (14.6645, 24.3307) and (14.5963, 24.3992) .. (14.5341, 24.4471).. controls (14.472, 24.495) and (14.4328, 24.5647) .. (14.3574, 24.5864).. controls (14.2394, 24.6204) and (14.1137, 24.6038) .. (13.991, 24.6089).. controls (13.8327, 24.6154) and (13.6754, 24.6587) .. (13.5178, 24.6427).. controls (13.4206, 24.6328) and (13.3262, 24.6005) .. (13.2286, 24.5969).. controls (13.182, 24.5951) and (13.1354, 24.6) .. (13.0888, 24.5986).. controls (13.0423, 24.5972) and (12.9944, 24.589) .. (12.9553, 24.5637).. controls (12.9205, 24.5413) and (12.895, 24.5069) .. (12.8752, 24.4705).. controls (12.8554, 24.4342) and (12.8407, 24.3953) .. (12.8225, 24.3581).. controls (12.7846, 24.2805) and (12.7322, 24.2109) .. (12.6853, 24.1383).. controls (12.6128, 24.0261) and (12.5527, 23.9056) .. (12.4708, 23.8).. controls (12.2783, 23.5517) and (11.9849, 23.4082) .. (11.7204, 23.2388).. controls (11.6632, 23.2021) and (11.6065, 23.1637) .. (11.5435, 23.1381).. controls (11.4806, 23.1126) and (11.4097, 23.1007) .. (11.3443, 23.1191).. controls (11.3347, 23.1219) and (11.3252, 23.1252) .. (11.3161, 23.1292) -- cycle;

  \path[draw=black,even odd rule,line cap=butt,line join=miter,line width=0.0141cm,dash pattern=on 0.0282cm off 0.0564cm] (6.2249, 23.7962) -- (5.1571, 23.7962);

  \path[draw=black,even odd rule,line cap=butt,line join=miter,line width=0.0176cm] (13.2617, 26.0926) -- (13.2617, 23.5871) -- (15.8123, 21.9881);

  \path[draw=black,even odd rule,line cap=butt,line join=miter,line width=0.0212cm] (6.3082, 24.5919) -- (6.3112, 23.7875) -- (7.1342, 23.7857);

  \path[draw=black,even odd rule,line cap=butt,line join=miter,line width=0.0176cm] (13.276, 23.5714) -- (16.0235, 24.3807);

  \path[draw=c242060,even odd rule,line cap=butt,line join=miter,line width=0.0529cm] (11.3443, 23.1191).. controls (11.5645, 23.1525) and (11.7194, 23.2231) .. (11.9088, 23.3401).. controls (12.6651, 23.8073) and (12.6419, 24.0739) .. (12.872, 24.5302).. controls (12.9501, 24.685) and (13.0242, 24.8443) .. (13.1364, 24.9764).. controls (13.1924, 25.0425) and (13.2579, 25.1013) .. (13.3331, 25.1444).. controls (13.4082, 25.1876) and (13.4933, 25.2148) .. (13.58, 25.2168).. controls (13.7137, 25.22) and (13.8434, 25.1634) .. (13.9503, 25.0831).. controls (14.0572, 25.0027) and (14.1444, 24.8995) .. (14.2295, 24.7963).. controls (14.3466, 24.6545) and (14.4622, 24.5106) .. (14.5969, 24.3855).. controls (14.7317, 24.2604) and (14.8876, 24.1539) .. (15.0635, 24.1003).. controls (15.2792, 24.0347) and (15.5194, 24.0541) .. (15.7217, 24.1533);

  \path[draw=c242060,even odd rule,line cap=butt,line join=miter,line width=0.0282cm,dash pattern=on 0.1693cm off 0.0564cm] (11.3382, 23.1693).. controls (11.5939, 23.2081) and (12.0367, 23.1952) .. (12.4534, 24.2668).. controls (12.6215, 24.6991) and (12.6293, 25.309) .. (12.8581, 25.3356).. controls (13.2423, 25.3802) and (13.2278, 24.5394) .. (13.6556, 24.6664).. controls (13.8233, 24.7162) and (14.0131, 25.0435) .. (14.4779, 24.4823).. controls (14.5441, 24.4024) and (14.8684, 24.1427) .. (15.1115, 24.0823).. controls (15.3962, 24.0116) and (15.6561, 24.1209) .. (15.7653, 24.1745);

  \node[text=black,line cap=butt,line join=miter,line width=0.0265cm,anchor=south west] (text120) at (5.246, 21.8555){$\omega\in\Omega$};

  \node[text=black,line cap=butt,line join=miter,line width=0.0265cm,anchor=south west] (text120-9-2) at (16.0559, 21.735){$y_1$};

  \node[text=black,line cap=butt,line join=miter,line width=0.0265cm,anchor=south west] (text120-9-2-2) at (13.3856, 23.0222){$c_1$};

  \node[text=black,line cap=butt,line join=miter,line width=0.0265cm,anchor=south west] (text120-9-2-6) at (16.2432, 24.3624){$y_2$};

  \node[text=black,line cap=butt,line join=miter,line width=0.0265cm,anchor=south west] (text120-9-2-6-9) at (12.8439, 26.2879){$f_{\boldsymbol{Y}}(\boldsymbol{y},t)$};

  \node[text=black,line cap=butt,line join=miter,line width=0.0265cm,rotate=0.0158,anchor=south west] (text120-9-7-6) at (8.3137, 21.4191){$Y^{1}_{t}=c_1$};

  \node[text=black,line cap=butt,line join=miter,line width=0.0265cm,rotate=0.0158,anchor=south west] (text120-9-7-6-3) at (3.7991, 25.2194){samples};

  \path[draw=black,even odd rule,line cap=butt,line join=miter,line width=0.0141cm,dash pattern=on 0.0282cm off 0.0564cm] (10.4749, 23.7713) -- (7.3031, 23.7713);

  \path[draw=black,even odd rule,line cap=butt,line join=miter,line width=0.0141cm,dash pattern=on 0.0282cm off 0.0564cm] (6.3119, 22.3735) -- (6.3119, 25.3085);

  \path[draw=black,even odd rule,line cap=butt,line join=miter,line width=0.0141cm,dash pattern=on 0.0282cm off 0.0564cm] (10.0444, 22.9231) -- (5.6016, 22.9231);

  \path[draw=black,even odd rule,line cap=butt,line join=miter,line width=0.0141cm,dash pattern=on 0.0282cm off 0.0564cm] (10.2715, 24.6439) -- (5.2488, 24.6439);

  \path[draw=black,even odd rule,line cap=butt,line join=miter,line width=0.0141cm,dash pattern=on 0.0282cm off 0.0564cm] (5.4642, 23.0795) -- (5.4642, 24.8569);

  \path[draw=black,even odd rule,line cap=butt,line join=miter,line width=0.0141cm,dash pattern=on 0.0282cm off 0.0564cm] (8.0642, 21.985) -- (8.0642, 25.1958);

  \path[draw=black,even odd rule,line cap=butt,line join=miter,line width=0.0141cm,dash pattern=on 0.0282cm off 0.0564cm] (7.2235, 22.0191) -- (7.2235, 25.2684);

  \path[draw=black,even odd rule,line cap=butt,line join=miter,line width=0.0141cm,dash pattern=on 0.0282cm off 0.0564cm] (8.9442, 22.1865) -- (8.9442, 25.2069);

  \path[draw=black,even odd rule,line cap=butt,line join=miter,line width=0.0141cm,dash pattern=on 0.0282cm off 0.0564cm] (9.8563, 22.6967) -- (9.8563, 24.9844);

  \node[text=black,line cap=butt,line join=miter,line width=0.0265cm,anchor=south west] (text120-9-7) at (5.752, 24.281){$X_2$};

  \node[text=black,line cap=butt,line join=miter,line width=0.0265cm,anchor=south west] (text120-9) at (6.7304, 23.3883){$X_1$};

  \path[draw=black,even odd rule,line cap=butt,line join=miter,line width=0.0176cm] (15.5687, 22.0405) -- (15.8123, 21.9881) -- (15.7051, 22.1365);

  \path[draw=black,even odd rule,line cap=butt,line join=miter,line width=0.0176cm] (15.817, 24.3892) -- (16.0235, 24.3807) -- (15.9193, 24.289);

  \path[draw=black,even odd rule,line cap=butt,line join=miter,line width=0.0176cm] (13.1826, 25.9174) -- (13.2617, 26.0926) -- (13.3318, 25.9257);

  \path[draw=black,even odd rule,line cap=butt,line join=miter,line width=0.0212cm] (6.2379, 24.4834) -- (6.3082, 24.6144) -- (6.3748, 24.4858);

  \path[draw=black,even odd rule,line cap=butt,line join=miter,line width=0.0212cm] (7.0178, 23.8423) -- (7.1576, 23.789) -- (7.0128, 23.7254);

  \path[draw=black,even odd rule,line cap=butt,line join=miter,line width=0.0176cm,dash pattern=on 0.1058cm off 0.1058cm] (11.713, 22.8226) -- (15.8535, 24.0781);

  \path[draw=black,line cap=round,line join=round,line width=0.0212cm] (7.5233, 21.0259) -- (7.5634, 21.8105);

  \path[draw=black,line cap=round,line join=round,line width=0.0212cm] (7.4555, 21.6065) -- (7.5634, 21.8105) -- (7.6429, 21.6081);

  \path[draw=black,line cap=round,line join=round,line width=0.0212cm] (7.4474, 21.4759) -- (7.5553, 21.6799) -- (7.6348, 21.4775);

  \path[draw=black,line cap=butt,line join=miter,line width=0.0212cm] (12.8653, 24.8424) -- (12.8449, 25.8524);

  \path[draw=black,line cap=butt,line join=miter,line width=0.0212cm] (12.7592, 25.5141) -- (12.8471, 25.7275) -- (12.9471, 25.515);

  \path[draw=black,line cap=butt,line join=miter,line width=0.0212cm] (12.7554, 25.6431) -- (12.8434, 25.8564) -- (12.9433, 25.644);

  \path[draw=black,line cap=butt,line join=miter,line width=0.0212cm] (13.5931, 25.7763) -- (13.5756, 24.7637);

  \path[draw=black,line cap=butt,line join=miter,line width=0.0212cm] (13.4869, 25.1046) -- (13.5749, 24.8913) -- (13.6748, 25.1037);

  \path[draw=black,line cap=butt,line join=miter,line width=0.0212cm] (13.486, 24.9628) -- (13.5739, 24.7495) -- (13.6739, 24.9619);

  \path[draw=black,line cap=round,line join=miter,line width=0.0212cm] (7.8881, 26.4595) -- (7.7104, 25.3589);

  \path[draw=black,line cap=round,line join=miter,line width=0.0212cm] (7.7682, 26.2858) -- (7.8914, 26.4745) -- (7.9627, 26.2531);

  \path[draw=black,line cap=round,line join=miter,line width=0.0212cm] (7.7407, 26.1498) -- (7.8639, 26.3385) -- (7.9352, 26.1171);

  \path[draw=black,line cap=butt,line join=miter,line width=0.0106cm] (8.2199, 21.4823).. controls (7.9382, 21.4684) and (7.6625, 21.7374) .. (7.6334, 21.9139);

  \path[draw=black,line cap=butt,line join=miter,line width=0.0106cm] (5.1963, 22.0512).. controls (5.0674, 23.0531) and (5.7636, 23.191) .. (5.8971, 23.2676);

  \node[text=black,line cap=butt,line join=miter,line width=0.0265cm,rotate=0.0158,anchor=south west] (text120-9-7-6-5) at (7.7136, 26.596){$\boldsymbol{U}\cdot\boldsymbol{n}$};

  \node[text=black,line cap=butt,line join=miter,line width=0.0265cm,rotate=0.0158,anchor=south west] (text120-9-7-6-5-9) at (7.2998, 20.6793){$\boldsymbol{U}\cdot\boldsymbol{n}$};

  \path[draw=black,fill=cd9aea4,fill opacity=0.2887,line cap=round,line join=round,line width=0.0141cm] (5.13, 24.9326) -- (5.361, 25.7038);

  \path[draw=black,fill=cd9aea4,fill opacity=0.2887,line cap=round,line join=round,line width=0.0141cm] (5.2617, 25.6259) -- (5.361, 25.7038);

  \path[draw=black,fill=cd9aea4,fill opacity=0.2887,line cap=round,line join=round,line width=0.0141cm] (5.4002, 25.5863) -- (5.361, 25.7038);

  \path[draw=c282463,fill opacity=0.3922,draw opacity=0.9451,line cap=butt,line join=miter,line width=0.0529cm] (7.6791, 25.2419).. controls (7.6791, 25.2419) and (7.6416, 24.9808) .. (7.6891, 24.767).. controls (7.7392, 24.5414) and (7.8752, 24.3541) .. (7.9011, 24.0095).. controls (7.9275, 23.6579) and (7.9026, 23.2109) .. (7.8125, 23.0462).. controls (7.5469, 22.5607) and (7.5594, 22.0066) .. (7.5594, 22.0066);

  \path[draw=c282463,fill opacity=0.3922,draw opacity=0.9451,line cap=butt,line join=miter,line width=0.0282cm] (8.3798, 25.2171).. controls (8.3798, 25.2171) and (8.3019, 24.8482) .. (8.3495, 24.6344).. controls (8.4813, 24.2268) and (8.8698, 23.788) .. (8.6859, 23.1338).. controls (8.5362, 22.601) and (8.6648, 22.1532) .. (8.6648, 22.1532);

  \path[draw=c282463,fill opacity=0.3922,draw opacity=0.9451,line cap=butt,line join=miter,line width=0.0282cm] (6.5911, 25.3363).. controls (6.5911, 25.3363) and (6.5851, 25.2885) .. (6.6326, 25.0747).. controls (6.6827, 24.8492) and (6.8187, 24.6619) .. (6.8446, 24.3173).. controls (6.871, 23.9656) and (6.6821, 23.5787) .. (6.5101, 23.3415).. controls (6.1853, 22.8934) and (5.8207, 22.6866) .. (5.8207, 22.6866);

  \path[draw=c282463,fill opacity=0.3922,draw opacity=0.9451,line cap=butt,line join=miter,line width=0.0282cm] (9.1783, 25.2077).. controls (9.1783, 25.2077) and (9.1407, 24.9466) .. (9.1882, 24.7328).. controls (9.2383, 24.5072) and (9.4986, 24.3855) .. (9.5553, 24.0445).. controls (9.611, 23.709) and (9.4965, 23.3824) .. (10.1756, 23.1004);

  \path[draw=c282463,fill opacity=0.3922,draw opacity=0.9451,line cap=butt,line join=miter,line width=0.0282cm] (5.5637, 25.0113).. controls (5.8406, 24.7372) and (5.6515, 24.6178) .. (5.6772, 24.3733).. controls (5.7142, 24.0226) and (5.5914, 23.6592) .. (5.5067, 23.4917).. controls (5.4, 23.2808) and (5.3075, 23.2064) .. (5.3075, 23.2064);

  \path[fill=black,draw opacity=0.9451,line cap=round,line join=round,line width=0.0282cm] (5.9754, 23.3081) circle (0.0441cm);

  \node[text=black,line cap=butt,line join=miter,line width=0.0265cm,anchor=south west] (text120-9-2-2-5) at (12.2615, 25.4674){in};

  \node[text=black,line cap=butt,line join=miter,line width=0.0265cm,anchor=south west] (text120-9-2-2-5-6) at (6.9995, 21.3206){in};

  \node[text=black,line cap=butt,line join=miter,line width=0.0265cm,anchor=south west] (text120-9-2-2-5-93) at (13.6961, 25.3745){out};

  \node[text=black,line cap=butt,line join=miter,line width=0.0265cm,anchor=south west] (text120-9-2-2-5-93-2) at (7.0284, 26.128){out};

  \path[draw=black,line cap=butt,line join=miter,line width=0.0106cm] (9.2444, 22.8009).. controls (9.3104, 22.4647) and (9.6869, 22.1734) .. (9.9696, 22.1496);

  \node[text=black,line cap=butt,line join=miter,line width=0.0265cm,rotate=0.0158,anchor=south west] (text120-9-7-6-9) at (10.1158, 21.9882){$\boldsymbol{Y}_{t}=\varphi_{t}(\boldsymbol{X})$};

\end{tikzpicture}

%% file: shapes.tex
\definecolor{c9f766c}{RGB}{159,118,108}
\definecolor{c53337b}{RGB}{83,51,123}
\definecolor{c492b7d}{RGB}{73,43,125}
\definecolor{ca37369}{RGB}{163,115,105}
\definecolor{c472b7b}{RGB}{71,43,123}

\def \globalscale {1.000000}
\begin{tikzpicture}[y=1cm, x=1cm, yscale=\globalscale,xscale=\globalscale, every node/.append style={scale=\globalscale}, inner sep=0pt, outer sep=0pt]
  \node[line width=0.02cm,anchor=south west] (text28) at (15.6049, 17.3189){$\frac{2}{\mu^{1}(\mathcal{X})}$};

  \path[draw=c9f766c,fill=c9f766c,fill opacity=0.3137,draw opacity=0.3137,line cap=butt,line join=miter,line width=0.0668cm] (16.1224, 20.6215) -- (16.1224, 18.5225);

  \path[fill=c53337b,fill opacity=0.8078,nonzero rule,draw opacity=0.3137,line cap=round,line join=round,line width=0.0525cm] (16.1205, 18.4873) circle (0.0978cm);

  \path[fill=c53337b,fill opacity=0.8078,nonzero rule,draw opacity=0.3137,line cap=round,line join=round,line width=0.0525cm] (16.125, 20.6313) circle (0.0978cm);

  \node[line width=0.02cm,anchor=south west] (text28-6-6-6-3-3-8) at (15.8621, 20.9973){$1$};

  \path[draw=c492b7d,fill=c9f766c,fill opacity=0.3137,line cap=round,line join=round,line width=0.0446cm] (13.2178, 20.3527) rectangle (14.7527, 18.8178);

  \node[line width=0.02cm,anchor=south west] (text28-6) at (13.3263, 17.3781){$\left({\frac{16}{\mu^{2}(\mathcal{X})}}\right)^{\frac{1}{2}}$};

  \node[line width=0.02cm,anchor=south west] (text28-6-6-6-3-3) at (13.7809, 20.9973){$2$};

  \path[draw=c492b7d,fill=ca37369,fill opacity=0.3137,line cap=round,line join=round,line width=0.0446cm] (11.1247, 19.5609) circle (1.07cm);

  \path[draw=c492b7d,fill=ca37369,fill opacity=0.3137,line cap=round,line join=round,line width=0.0446cm] (8.0562, 19.5609) circle (1.07cm);

  \node[line width=0.02cm,anchor=south west] (text28-6-6) at (10.3303, 17.3251){$\left({\frac{4\pi}{\mu^{2}(\mathcal{X})}}\right)^{\frac{1}{2}}$};

  \node[line width=0.02cm,anchor=south west] (text28-6-6-6-3-5) at (10.929, 20.9973){$2$};

  \path[fill=white,fill opacity=0.4393,line cap=round,line join=round,line width=0.0393cm] (8.0575, 19.5604) circle (0.9428cm);

  \path[fill=white,fill opacity=0.6107,line cap=round,line join=round,line width=0.0302cm] (8.0505, 19.5669) circle (0.7242cm);

  \path[fill=white,fill opacity=0.9262,line cap=round,line join=round,line width=0.0156cm] (8.055, 19.5608) circle (0.3748cm);

  \path[draw=c492b7d,fill=ca37369,fill opacity=0.3137,line cap=round,line join=round,line width=0.0446cm] (4.8007, 19.5537) circle (1.07cm);

  \path[fill=white,fill opacity=0.4393,line cap=round,line join=round,line width=0.0393cm] (4.8022, 19.563) circle (0.9428cm);

  \path[fill=white,fill opacity=0.6107,line cap=round,line join=round,line width=0.0302cm] (4.7952, 19.5695) circle (0.7242cm);

  \path[fill=white,fill opacity=0.9262,line cap=round,line join=round,line width=0.0156cm] (4.7997, 19.5634) circle (0.3748cm);

  \node[line width=0.02cm,anchor=south west] (text28-6-6-6) at (7.3142, 17.3251){$\left(\frac{36\pi}{\mu^{3}(\mathcal{X})}\right)^{\frac{1}{3}}$};

  \node[line width=0.02cm,anchor=south west] (text28-6-6-6-3) at (7.7837, 20.9973){$3$};

  \node[line width=0.02cm,anchor=south west] (text28-6-6-6-3-59) at (4.5502, 20.9934){$d$};

  \path[draw=c472b7b,line cap=round,line join=round,line width=0.0213cm,dash pattern=on 0.1707cm off 0.0854cm] (4.8007, 19.5582) circle (1.2686cm);

  \path[draw=c472b7b,line cap=round,line join=round,line width=0.0213cm,dash pattern=on 0.1707cm off 0.0854cm] (4.8016, 19.5647) circle (0.4132cm);

  \node[line width=0.02cm,anchor=south west] (text28-6-6-6-4) at (3.3479, 17.1601){$\left(\frac{2^{d}\pi^{\frac{d}{2}}\Gamma\left(\frac{d}{2}+1\right)^{1-d}}{\Gamma\left(\frac{d}{2}\right)^{d}\mu^{d}(\mathcal{X})}\right)^{\frac{1}{d}}$};

\end{tikzpicture}

%% file: diag_D2.tex
\definecolor{ce4e4e4}{RGB}{228,228,228}
\definecolor{c1f77b4}{RGB}{31,119,180}
\definecolor{cd80000}{RGB}{216,0,0}

\def \globalscale {1.000000}
\begin{tikzpicture}[y=1cm, x=1cm, yscale=\globalscale,xscale=\globalscale, every node/.append style={scale=\globalscale}, inner sep=0pt, outer sep=0pt]
  \path[fill=ce4e4e4,line cap=round,line join=round,line width=0.016cm] (8.2155, 23.7574) -- (6.9506, 24.0975).. controls (6.9506, 24.0975) and (6.8943, 24.0926) .. (6.8408, 23.8388).. controls (6.7966, 23.6288) and (6.932, 23.5821) .. (6.932, 23.5821) -- cycle;

  \path[fill=ce4e4e4,line cap=round,line join=round,line width=0.016cm] (8.225, 23.7601) -- (11.9639, 22.6869).. controls (11.9639, 22.6869) and (12.1774, 23.0992) .. (12.0734, 23.444).. controls (11.9142, 23.9724) and (12.1535, 24.3267) .. (12.1535, 24.3267) -- cycle;

  \path[draw=c1f77b4,line cap=round,line join=round,line width=0.0321cm] (11.8961, 22.7232) -- (7.0509, 24.0681);

  \path[draw=black,line cap=round,line join=round,line width=0.0257cm] (8.214, 23.7504) -- (8.906, 26.2693);

  \path[draw=black,line cap=round,line join=round,line width=0.0257cm] (8.6913, 25.9692) -- (8.9104, 26.2838) -- (8.9666, 25.906);

  \path[draw=c1f77b4,line cap=round,line join=round,line width=0.0321cm] (12.0125, 24.3109) -- (7.0271, 23.5967);

  \path[draw=black,line cap=round,line join=round,line width=0.0257cm] (8.2123, 23.7514) -- (7.801, 26.6659);

  \path[draw=black,line cap=round,line join=round,line width=0.0257cm] (7.6872, 26.266) -- (7.7957, 26.6774) -- (8.0155, 26.3228);

  \path[draw=black,line cap=round,line join=round,line width=0.016cm] (7.9978, 23.813) -- (8.0603, 24.0461) -- (8.273, 23.9891);

  \path[draw=black,line cap=round,line join=round,line width=0.016cm] (7.9913, 23.7203) -- (7.9583, 23.9575) -- (8.1741, 23.987);

  \path[draw=cd80000,line cap=round,line join=round,line width=0.016cm] (8.2208, 23.7585) -- (9.8298, 24.9331);

  \path[draw=cd80000,line cap=round,line join=round,line width=0.016cm] (9.6091, 24.8867) -- (9.8289, 24.936) -- (9.7408, 24.7212);

  \path[draw=cd80000,line cap=round,line join=round,line width=0.016cm] (9.4873, 24.8055) -- (9.7072, 24.8547) -- (9.619, 24.6399);

  \node[line cap=round,line join=round,line width=0.016cm,anchor=south west] (text14) at (9.8657, 24.9997){$\boldsymbol{u}$};

  \node[line cap=round,line join=round,line width=0.016cm,anchor=south west] (text14-6) at (7.5287, 26.8431){$\vc{v}$};

  \node[line cap=round,line join=round,line width=0.016cm,anchor=south west] (text14-6-0) at (9.4621, 23.563){$\boldsymbol{u}^{\top}\boldsymbol{\alpha}\boldsymbol{v}\boldsymbol{v}^{\top}\boldsymbol{u}<0$};

  \node[line cap=round,line join=round,line width=0.016cm,anchor=south west] (text14-6-0-9) at (9.4904, 25.9383){$\boldsymbol{u}^{\top}\boldsymbol{\alpha}\boldsymbol{v}\boldsymbol{v}^{\top}\boldsymbol{u}>0$};

  \node[line cap=round,line join=round,line width=0.016cm,anchor=south west] (text14-6-1) at (8.7896, 26.4425){$\boldsymbol{\alpha}\boldsymbol{v}$};

  \path[draw=black,line cap=round,line join=round,line width=0.0128cm] (8.0294, 25.5016).. controls (8.0294, 25.5016) and (8.2817, 25.5596) .. (8.5862, 25.4603);

  \path[draw=black,line cap=round,line join=round,line width=0.0128cm] (8.521, 25.5438) -- (8.6002, 25.454) -- (8.496, 25.414);

\end{tikzpicture}

%% file: boxes.pdf_tex
\begingroup%
  \makeatletter%
  \providecommand\color[2][]{%
    \errmessage{(Inkscape) Color is used for the text in Inkscape, but the package 'color.sty' is not loaded}%
    \renewcommand\color[2][]{}%
  }%
  \providecommand\transparent[1]{%
    \errmessage{(Inkscape) Transparency is used (non-zero) for the text in Inkscape, but the package 'transparent.sty' is not loaded}%
    \renewcommand\transparent[1]{}%
  }%
  \providecommand\rotatebox[2]{#2}%
  \newcommand*\fsize{\dimexpr\f@size pt\relax}%
  \newcommand*\lineheight[1]{\fontsize{\fsize}{#1\fsize}\selectfont}%
  \ifx\svgwidth\undefined%
    \setlength{\unitlength}{421.65561946bp}%
    \ifx\svgscale\undefined%
      \relax%
    \else%
      \setlength{\unitlength}{\unitlength * \real{\svgscale}}%
    \fi%
  \else%
    \setlength{\unitlength}{\svgwidth}%
  \fi%
  \global\let\svgwidth\undefined%
  \global\let\svgscale\undefined%
  \makeatother%
  \begin{picture}(1,0.42035746)%
    \lineheight{1}%
    \setlength\tabcolsep{0pt}%
    \put(0,0){\includegraphics[width=\unitlength,page=1]{boxes.pdf}}%
    \put(0.0427575,0.37513357){\makebox(0,0)[lt]{\lineheight{1.25}\smash{\begin{tabular}[t]{l}$(a)$\end{tabular}}}}%
    \put(0.47722189,0.3757638){\makebox(0,0)[lt]{\lineheight{1.25}\smash{\begin{tabular}[t]{l}$(b)$\end{tabular}}}}%
    \put(0.11412159,0.03479085){\makebox(0,0)[lt]{\lineheight{1.25}\smash{\begin{tabular}[t]{l}$\theta_1$\end{tabular}}}}%
    \put(0.31536982,0.07371904){\makebox(0,0)[lt]{\lineheight{1.25}\smash{\begin{tabular}[t]{l}$\theta_2$\end{tabular}}}}%
    \put(-0.00383281,0.2043326){\makebox(0,0)[lt]{\lineheight{1.25}\smash{\begin{tabular}[t]{l}$\theta_3$\end{tabular}}}}%
    \put(0.55242494,0.03555271){\makebox(0,0)[lt]{\lineheight{1.25}\smash{\begin{tabular}[t]{l}$\theta_1$\end{tabular}}}}%
    \put(0.75367127,0.074481){\makebox(0,0)[lt]{\lineheight{1.25}\smash{\begin{tabular}[t]{l}$\theta_2$\end{tabular}}}}%
    \put(0.43447057,0.20509456){\makebox(0,0)[lt]{\lineheight{1.25}\smash{\begin{tabular}[t]{l}$\theta_3$\end{tabular}}}}%
  \end{picture}%
\endgroup%

%% file: diag_leibniz.tex
\definecolor{c1a028b}{RGB}{26,2,139}
\definecolor{c7e7e7e}{RGB}{126,126,126}
\definecolor{cacacac}{RGB}{172,172,172}
\definecolor{cb90000}{RGB}{185,0,0}

\def \globalscale {1.000000}
\begin{tikzpicture}[y=1cm, x=1cm, yscale=\globalscale,xscale=\globalscale, every node/.append style={scale=\globalscale}, inner sep=0pt, outer sep=0pt]
  \path[draw=c1a028b,fill=c7e7e7e,fill opacity=0.1402,line cap=butt,line join=miter,line width=0.0304cm] (6.2083, 25.7314).. controls (6.2083, 26.143) and (6.438, 26.5335) .. (6.7945, 26.7392).. controls (7.1509, 26.945) and (7.5901, 26.945) .. (7.9465, 26.7392).. controls (8.303, 26.5335) and (8.5226, 26.1531) .. (8.5226, 25.7415) -- (8.5124, 24.4481).. controls (8.5124, 24.0366) and (8.2928, 23.6562) .. (7.9364, 23.4504).. controls (7.5799, 23.2446) and (7.1408, 23.2446) .. (6.7843, 23.4504).. controls (6.4279, 23.6562) and (6.2083, 24.0366) .. (6.2083, 24.4481) -- cycle;

  \path[draw=black,line cap=butt,line join=miter,line width=0.0142cm] (9.7341, 27.3316) -- (9.7341, 22.8841) -- (11.8633, 22.8841);

  \path[draw=black,line cap=butt,line join=miter,line width=0.0142cm] (11.7412, 22.7944) -- (11.913, 22.8851) -- (11.7296, 22.9833);

  \path[draw=black,line cap=butt,line join=miter,line width=0.0142cm] (9.6335, 27.1416) -- (9.7341, 27.3483) -- (9.8202, 27.1297);

  \path[draw=black,line cap=butt,line join=miter,line width=0.0142cm] (12.4492, 27.3186) -- (12.4492, 22.8841) -- (15.0529, 22.8841);

  \path[draw=black,line cap=butt,line join=miter,line width=0.0142cm] (14.8895, 22.79) -- (15.0613, 22.8807) -- (14.8779, 22.9788);

  \path[draw=black,line cap=butt,line join=miter,line width=0.0142cm] (12.3486, 27.1416) -- (12.4492, 27.3483) -- (12.5353, 27.1297);

  \path[draw=black,fill=cacacac,fill opacity=0.1402,line cap=round,line join=round,line width=0.0149cm,dash pattern=on 0.1195cm off 0.0598cm] (7.3622, 25.6927) ellipse (1.1435cm and 0.3812cm);

  \path[draw=black,fill=c7e7e7e,fill opacity=0.1402,line cap=round,line join=round,line width=0.0108cm] (7.3695, 23.7232) ellipse (0.847cm and 0.2671cm);

  \path[draw=black,line cap=round,line join=round,line width=0.017cm] (6.4867, 23.6721) -- (5.2683, 22.4654);

  \path[draw=black,line cap=round,line join=round,line width=0.0113cm,dash pattern=on 0.0907cm off 0.0907cm] (5.2456, 22.4441) -- (4.8505, 22.0469);

  \path[draw=black,line cap=round,line join=round,line width=0.017cm] (5.3141, 22.6477) -- (5.2627, 22.4636) -- (5.4416, 22.5029);

  \path[draw=black,line cap=round,line join=round,line width=0.0113cm] (4.8959, 22.2316) -- (4.8427, 22.0413) -- (5.0276, 22.0818);

  \path[draw=black,line cap=round,line join=round,line width=0.017cm] (6.4671, 23.6789) -- (4.7707, 23.6799);

  \path[draw=black,line cap=round,line join=round,line width=0.0085cm,dash pattern=on 0.034cm off 0.068cm] (7.4037, 23.6871) -- (6.5036, 23.6881);

  \path[draw=black,line cap=round,line join=round,line width=0.0085cm,dash pattern=on 0.034cm off 0.068cm] (6.4898, 23.063) -- (6.4909, 23.9631);

  \path[draw=black,line cap=round,line join=round,line width=0.0085cm,dash pattern=on 0.034cm off 0.068cm] (13.953, 22.6892) -- (13.9541, 24.3624);

  \path[draw=black,line cap=round,line join=round,line width=0.0085cm,dash pattern=on 0.034cm off 0.068cm] (14.2551, 24.48) -- (13.6827, 24.481);

  \path[draw=black,line cap=round,line join=round,line width=0.0085cm,dash pattern=on 0.034cm off 0.068cm] (10.9626, 22.7048) -- (10.9636, 23.2006);

  \path[draw=black,line cap=round,line join=round,line width=0.017cm] (4.9171, 23.7758) -- (4.7556, 23.6833) -- (4.9064, 23.5883);

  \path[draw=black,line cap=butt,line join=miter,line width=0.0113cm] (4.7114, 24.3555) -- (4.7114, 24.0233);

  \path[draw=black,line cap=butt,line join=miter,line width=0.0113cm] (4.7101, 24.1963) -- (6.4954, 24.1963);

  \path[draw=black,line cap=butt,line join=miter,line width=0.0113cm] (6.4952, 24.3607) -- (6.4952, 24.0344);

  \path[draw=black,line cap=butt,line join=miter,line width=0.0113cm] (4.296, 22.0191) -- (4.6282, 22.0191);

  \path[draw=black,line cap=butt,line join=miter,line width=0.0113cm] (4.4552, 22.0125) -- (4.4552, 23.6644);

  \path[draw=black,line cap=butt,line join=miter,line width=0.0113cm] (4.2907, 23.6633) -- (4.617, 23.6633);

  \path[draw=black,line cap=butt,line join=miter,line width=0.0113cm] (8.7729, 23.3011) -- (9.1051, 23.3011);

  \path[draw=black,line cap=butt,line join=miter,line width=0.0113cm] (12.3354, 23.283) -- (12.4969, 23.283);

  \path[draw=black,line cap=butt,line join=miter,line width=0.0113cm] (12.2794, 24.5109) -- (12.4409, 24.5109);

  \path[draw=black,line cap=butt,line join=miter,line width=0.0113cm] (12.2822, 25.6966) -- (12.4437, 25.6966);

  \path[draw=black,line cap=butt,line join=miter,line width=0.0113cm] (12.2884, 26.8858) -- (12.4499, 26.8858);

  \path[draw=black,line cap=butt,line join=miter,line width=0.0113cm] (8.9321, 23.2945) -- (8.9321, 26.8793);

  \path[draw=black,line cap=butt,line join=miter,line width=0.0113cm] (8.7677, 24.4859) -- (9.094, 24.4859);

  \path[draw=black,line cap=butt,line join=miter,line width=0.0113cm] (8.7677, 25.6899) -- (9.094, 25.6899);

  \path[draw=black,line cap=butt,line join=miter,line width=0.0113cm] (8.7677, 26.8948) -- (9.094, 26.8948);

  \path[draw=black,line cap=butt,line join=miter,line width=0.0113cm] (5.7997, 23.6703).. controls (5.7997, 23.6703) and (5.8014, 23.5698) .. (5.8267, 23.4667).. controls (5.8465, 23.386) and (5.8948, 23.2694) .. (5.9813, 23.1749);

  \path[draw=black,fill=white,fill opacity=0.6621,line cap=round,line join=round,line width=0.0079cm] (6.49, 23.7831) rectangle (6.6358, 23.6841);

  \path[draw=black,fill=white,fill opacity=0.6621,line cap=round,line join=round,line width=0.0095cm,rotate around={-44.5898:(0.0, 29.7)}] (8.8555, 29.9634) rectangle (8.9998, 29.8192);

  \path[draw=cb90000,line cap=round,line join=round,line width=0.0425cm] (10.9571, 23.2907) -- (10.9571, 26.8947);

  \path[draw=cb90000,line cap=round,line join=round,line width=0.0412cm] (13.9426, 24.4863) -- (13.9426, 25.7109);

  \path[draw=cb90000,line cap=round,line join=round,line width=0.0412cm] (12.4439, 26.8871).. controls (13.0039, 26.8093) and (13.928, 26.3886) .. (13.9385, 25.696);

  \path[draw=cb90000,line cap=round,line join=round,line width=0.0412cm] (12.439, 23.2876).. controls (12.9989, 23.3653) and (13.923, 23.7861) .. (13.9335, 24.4786);

  \path[draw=cb90000,line cap=round,line join=round,line width=0.0213cm] (10.9571, 26.8947) -- (9.7341, 26.8947);

  \path[draw=cb90000,line cap=round,line join=round,line width=0.0213cm] (10.9571, 23.2907) -- (9.7341, 23.2907);

  \node[line width=0.0213cm,anchor=south west] (text39) at (10.8438, 22.071){$\frac{1}{3}$};

  \node[line width=0.0213cm,anchor=south west] (text39-3) at (9.309, 27.233){$z$};

  \node[line width=0.0213cm,anchor=south west] (text39-3-8) at (5.3881, 24.3249){$1$};

  \node[line width=0.0213cm,anchor=south west] (text39-3-8-4) at (5.5378, 23.2742){$\gamma$};

  \node[line width=0.0213cm,anchor=south west] (text39-3-8-4-6) at (5.9578, 22.8772){$\boldsymbol{n}$};

  \node[line width=0.0213cm,anchor=south west] (text39-3-8-4-6-9) at (5.4813, 21.6504){$\dfrac{\boldsymbol{n}}{\mathrm{cos}(\gamma)}$};

  \node[line width=0.0213cm,anchor=south west] (text39-3-8-4-6-9-8) at (3.2872, 22.679){$\mathrm{tan}(\gamma)$};

  \node[line width=0.0213cm,anchor=south west] (text39-3-1) at (12.0077, 27.2366){$z$};

  \node[line width=0.0213cm,anchor=south west] (text39-2) at (11.4433, 22.272){$f_{Z|\partial\Omega}$};

  \node[line width=0.0213cm,anchor=south west] (text39-2-8) at (14.8077, 22.3387){$f_{Z}$};

  \node[line width=0.0213cm,anchor=south west] (text39-5) at (9.0845, 23.8542){$1$};

  \node[line width=0.0213cm,anchor=south west] (text39-5-1) at (11.9538, 23.2102){$0$};

  \node[line width=0.0213cm,anchor=south west] (text39-5-1-2) at (11.9475, 26.7999){$3$};

  \node[line width=0.0213cm,anchor=south west] (text39-5-1-2-9) at (11.965, 25.6295){$2$};

  \node[line width=0.0213cm,anchor=south west] (text39-5-1-2-9-2) at (11.9917, 24.4471){$1$};

  \node[line width=0.0213cm,anchor=south west] (text39-5-7) at (9.1027, 24.986){$1$};

  \node[line width=0.0213cm,anchor=south west] (text39-5-7-3) at (9.1113, 26.1749){$1$};

  \node[line width=0.0213cm,anchor=south west] (text39-1) at (13.7984, 22.0581){$\frac{3}{7}$};

  \node[line width=0.0213cm,anchor=south west] (text39-1-8) at (14.4497, 23.3637){$\frac{3}{7}(2z-z^{2})$};

  \path[draw=black,line cap=butt,line join=miter,line width=0.0113cm] (13.6576, 23.7427).. controls (13.6576, 23.7427) and (13.8905, 23.5494) .. (14.2936, 23.5479);

\end{tikzpicture}